\documentclass[aps,prl,reprint,superscriptaddress,nofootinbib,floatfix]{revtex4-2}

\usepackage[T1]{fontenc}
\usepackage[utf8]{inputenc}
\usepackage{amsmath,amssymb}
\usepackage{graphicx}
\usepackage{bm}
\usepackage{microtype}
\usepackage[hidelinks]{hyperref}
\usepackage{xcolor}

\ifdefined\reviewcopy
  \usepackage[switch]{lineno}
\fi

\graphicspath{{./}}
\begin{document}

\title{Power–Bandwidth Scaling of Resonantly Coupled Soliton Microcombs}

\author{Xinrui Luo$^{1,*}$, Kaixuan Zhu$^{1,*}$, Yuanlei Wang$^{1,2*}$, Yinke Cheng$^{1,2}$, Haoyang Luo$^{1}$, Junqi Wang$^{1}$, Yiwen Yang$^{1}$, Zhenyu Xie$^{1}$, Bei-Bei Li$^{2}$, Qihuang Gong$^{1,3,4,5}$, and Qi-Fan Yang$^{1,3,4,5\dagger}$\\
$^1$State Key Laboratory for Artificial Microstructure and Mesoscopic Physics and Frontiers Science Center for Nano-optoelectronics, School of Physics, Peking University, Beijing 100871, China\\
$^2$Beijing National Laboratory for Condensed Matter Physics, Institute of Physics, Chinese Academy of Sciences, Beijing 100190, China\\
$^3$Peking University Yangtze Delta Institute of Optoelectronics, Nantong 226010, China\\
$^4$Collaborative Innovation Center of Extreme Optics, Shanxi University, Taiyuan 030006, China\\
$^5$Hefei National Laboratory, Hefei 230088, China\\
$^{*}$These authors contributed equally to this work.\\
$^{\dagger}$Corresponding author: leonardoyoung@pku.edu.cn}

\begin{abstract}
A soliton microcomb requires increasing pump power as its optical bandwidth is broadened. Resonant pumping through an auxiliary microresonator can reduce the power required to sustain a soliton, but simultaneously increases the power required for soliton formation. We show that this competition leads to optimal inter-resonator coupling, and the predicted minimum input pump power scales as the two-thirds power of the comb bandwidth, in contrast to the quadratic scaling under direct pumping. Experiments support the opposing power trends, providing a design rule for power-efficient, ultra-broadband soliton microcombs.
\end{abstract}

\maketitle

\ifdefined\reviewcopy
  \linenumbers
\fi

\textit{Introduction.---} A dissipative Kerr soliton is a localized pulse sustained by the balance between dispersion and Kerr nonlinearity in a continuously pumped microresonator \cite{herr2014temporal,kippenberg2018dissipative}. Its periodic circulation generates a coherent optical frequency comb---a soliton microcomb---with a line spacing set by the resonator round-trip rate \cite{herr2014temporal,kippenberg2018dissipative,diddams2020optical,delhaye2007optical,kippenberg2011microresonator}. Soliton microcombs have enabled molecular spectroscopy \cite{suh2016microresonator,dutt2018chip,yang2019vernier}, optical communications \cite{marinpalomo2017microresonator,rizzo2022petabit,jrgensen2022petabit}, and precision frequency measurement \cite{spencer2018optical,newman2019architecture}. Broader spectra are desirable because they provide more coherent comb lines and extend the accessible optical bandwidth for these applications. In the conventional architecture, the microresonator is pumped directly through a bus waveguide, for which the input power required to sustain a soliton scales quadratically with the comb bandwidth \cite{lucas2017detuning,li2018universal}. This scaling can place ultra-broadband soliton microcombs beyond the power available from integrated pump lasers \cite{stern2018battery,raja2019electrically,shen2020integrated,xiang2021laser}.

Several approaches have been developed to reduce the pump-power requirement for broadband soliton microcombs \cite{yang2024efficient}. One approach introduces an auxiliary microresonator, referred to as the resonant coupler (RC), between the bus waveguide and the comb-forming nonlinear resonator (NR), thereby resonantly enhancing the pump power delivered to the NR \cite{xue2019super,zhu2026power}. This enhancement can extend the accessible comb bandwidth at a given input power, providing a route toward octave-spanning soliton microcombs at microwave repetition rates for direct optical-to-microwave links.

The benefit of using an RC to generate solitons from a continuous-wave state must also account for the soliton-formation process. To form solitons, the pump laser is typically tuned from the blue to the red side of the resonance, during which the continuous-wave state develops modulation instability (MI), whose sidebands can seed soliton formation \cite{herr2014temporal}. Further increasing the detuning after formation broadens the soliton spectrum. These two stages, referred to as soliton formation and sustainment, generally impose different pump-power requirements. For a fixed laser input power, the minimum required power is therefore determined by the larger of the formation and sustainment requirements. In this Letter, we show that RC--NR coupling affects these two requirements in opposite ways, giving rise to an optimal coupling at which they are balanced. At this optimum, the minimum required power asymptotically scales as the two-thirds power of the soliton-microcomb bandwidth.

\begin{figure*}[!t]
\centering
\includegraphics[width=\textwidth]{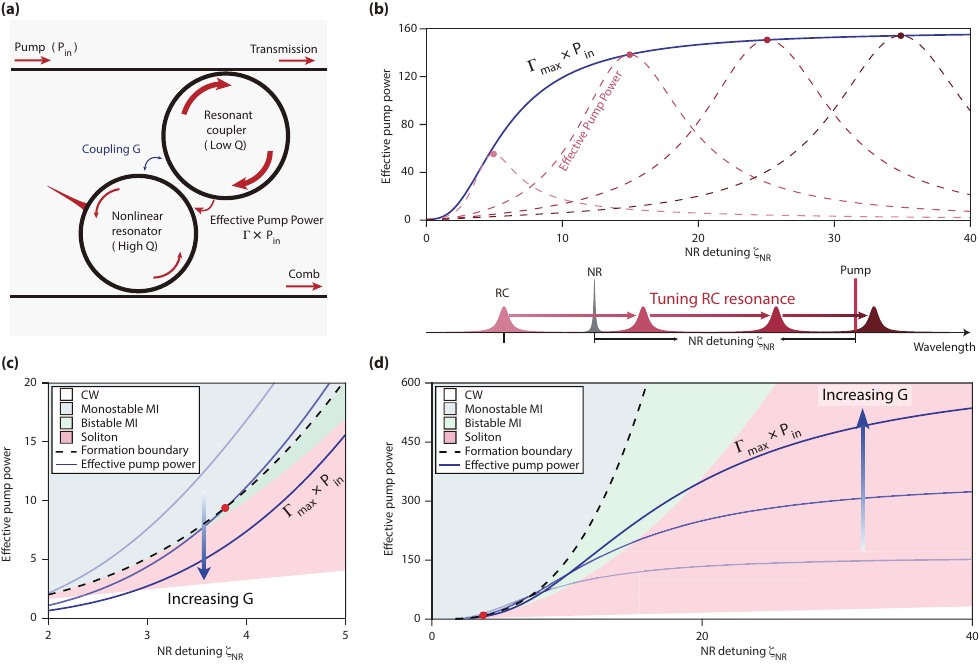}
\caption{\label{fig:effective-pump}
\textbf{Effective pump power in resonantly coupled microresonators.}
\textbf{(a)} Coupled-resonator configuration.
\textbf{(b)} Effective pump power $|f_\mathrm{eff}|^2$ versus NR detuning for different preset RC resonance frequencies. The solid curve gives the maximum over all RC settings. The lower schematic illustrates the tuning of the RC resonance relative to the NR and pump.
\textbf{(c,d)} Effective pump power overlaid on the NR phase diagram at small and large detunings, respectively.}
\end{figure*}

\textit{Theoretical model.---} The configuration is shown in Fig.~\ref{fig:effective-pump}(a). A continuous-wave pump with power $P_\mathrm{in}$ is coupled from the bus waveguide into the RC, which is in turn coupled to the NR at a rate $G$. The total decay rates of the NR and RC are denoted by $\kappa_\mathrm{NR(RC)}$, with corresponding external coupling rates $\kappa_{e,\mathrm{NR(RC)}}$. The comb generated in the NR is collected from the drop port.

We first adopt a reduced model in which the RC provides an effective drive to the pump mode of the NR. The assumptions underlying this reduction and their relation to the full coupled-LLE simulations are detailed in the Supplementary Materials. The intracavity dynamics of the NR are described by the normalized Lugiato--Lefever equation,
\begin{equation}
\frac{\partial\psi_\mathrm{NR}}{\partial\tau}
=-(1+i\zeta_\mathrm{NR})\psi_\mathrm{NR}
+id_2\frac{\partial^2\psi_\mathrm{NR}}{\partial\phi^2}
+i|\psi_\mathrm{NR}|^2\psi_\mathrm{NR}
+f_\mathrm{eff}.
\label{eq}
\end{equation}
Here, $\psi_\mathrm{NR}$ is the normalized intracavity field, $\tau=\kappa_\mathrm{NR}t/2$ is the normalized slow time, and $\phi$ is the azimuthal coordinate. The normalized pump detunings are defined as $\zeta_\mathrm{NR(RC)}=2(\omega_\mathrm{NR(RC)}-\omega_p)/\kappa_\mathrm{NR}$, where $\omega_\mathrm{NR}$, $\omega_\mathrm{RC}$, and $\omega_p$ denote the NR resonance, RC resonance, and pump angular frequencies, respectively. The normalized second-order dispersion is denoted by $d_2$.

\begin{figure}[t]
\centering
\includegraphics[width=\linewidth]{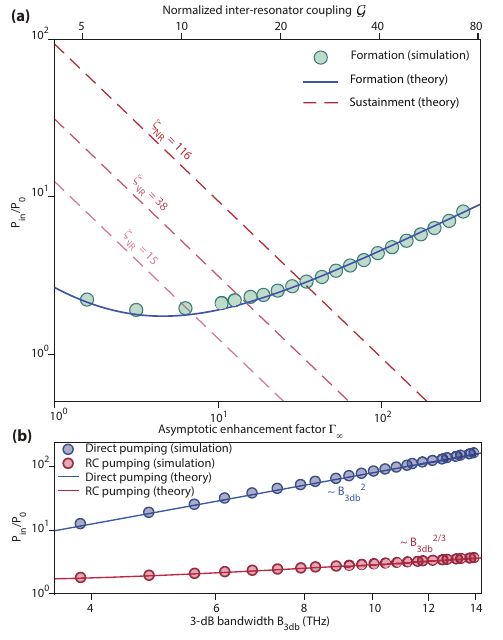}
%width=\textwidth，width=\linewidth
\caption{\label{fig:design}
\textbf{Formation, sustainment, and power--bandwidth scaling.}
\textbf{(a)} Normalized pump power versus the asymptotic enhancement factor $\Gamma_\infty$, with the normalized inter-resonator coupling $\mathcal G$ shown on the upper axis. The blue curve shows the analytical formation power given by Eq.~\eqref{eq:formation-input}, with green circles indicating formation powers obtained from simulations. Red dashed curves show the soliton sustainment power Eq.~\eqref{eq:sustainment-input} at the indicated NR detunings.
\textbf{(b)} Minimum pump power versus 3-dB soliton bandwidth for pumping through RC (red) and direct pumping through bus waveguide (blue). Solid curves show analytical predictions, and circles denote simulations. For RC, the inter-resonator coupling is optimized for each target bandwidth.}
\end{figure}

The key quantity considered in this work is the normalized effective pump power, $|f_\mathrm{eff}|^2$, delivered from the RC to the NR. To relate this quantity to the input power, we first evaluate the transfer of the pump field through the coupled resonators using their linear cavity response while retaining the Kerr nonlinearity in the subsequent NR dynamics. Under this approximation, $|f_\mathrm{eff}|^2$ is related to the bus-waveguide input power, $P_\mathrm{in}$, by
\begin{equation}
|f_\mathrm{eff}|^2
=\frac{\Gamma P_\mathrm{in}}
{P_0},
\end{equation}
where $P_0=\hbar\omega_0\kappa_\mathrm{NR}^3/(8g_\mathrm{NR}\kappa_{e,\mathrm{NR}})$, with $\hbar$ the reduced Planck constant, $\omega_0$ the optical angular frequency, and $g_\mathrm{NR}$ the Kerr nonlinear coupling coefficient of the NR. Notably, $P_0$ also corresponds to the parametric oscillation threshold of the NR when directly pumped through its bus waveguide. Relative to direct waveguide pumping, the RC resonantly enhances the pump power delivered to the NR by a frequency-dependent factor $\Gamma$,
\begin{equation}
\Gamma(\zeta_\mathrm{NR},\zeta_\mathrm{RC})
=
\frac{K_e\mathcal G^2}
{\left(K+\dfrac{\mathcal G^2}{1+\zeta_\mathrm{NR}^2}\right)^2
+\left(\zeta_\mathrm{RC}
-\dfrac{\mathcal G^2\zeta_\mathrm{NR}}
{1+\zeta_\mathrm{NR}^2}\right)^2}.
\end{equation}
Here, $\mathcal G=2G/\kappa_\mathrm{NR}$, $K=\kappa_\mathrm{RC}/\kappa_\mathrm{NR}$, and $K_e=\kappa_{e,\mathrm{RC}}/\kappa_{e,\mathrm{NR}}$. This expression follows from the linear response of the coupled resonators, but we use it here to describe the effective pump transfer in the nonlinear system under the conditions validated by the full coupled-LLE simulations in the Supplementary Materials. For a target NR detuning $\zeta_\mathrm{NR}$, the enhancement is maximized by choosing
$\zeta_\mathrm{RC}=\mathcal G^2\zeta_\mathrm{NR}/(1+\zeta_\mathrm{NR}^2)$, yielding
\begin{equation}
\Gamma_\mathrm{max}(\zeta_\mathrm{NR})
=\frac{K_e\mathcal G^2}
{\left[K+\mathcal G^2/(1+\zeta_\mathrm{NR}^2)\right]^2}.
\label{eq:maximum-enhancement}
\end{equation}
The corresponding maximum effective pump power is plotted as a function of NR detuning in Fig.~\ref{fig:effective-pump}(b) and increases with $\zeta_\mathrm{NR}$. At sufficiently large detuning, the enhancement approaches the asymptotic value $\Gamma_\infty=K_e\mathcal G^2/K^2$ (see Supplementary Materials). The optimal route to a target soliton, defined as the trajectory requiring the minimum input pump power, therefore follows this maximum-enhancement condition by dynamically adjusting the RC resonance together with the pump-laser tuning.

\textit{Pump power requirements.---} We next overlay the optimal route on the NR phase diagram. The MI region exhibits fine structure [Fig.~\ref{fig:effective-pump}(c)], and the portion continuously accessible from the continuous-wave state is referred to as monostable MI. Soliton formation therefore requires the optimal route to cross the right boundary of the monostable-MI region (denoted as the formation boundary). This crossing occurs at relatively small NR detuning, where increasing $G$ can reduce the effective pump power and thereby raises the required input pump power to have such a crossing. Such formation power shall satisfy (see Supplementary Materials)
\begin{equation}
\begin{aligned}
P_\mathrm{in}&\geq
\frac{8\sqrt3}{9K_e}
\frac{\sqrt K\left(K+\mathcal G^2\right)^{3/2}}
{\mathcal G^2}P_0 \\
&=\frac{8\sqrt3}{9}
\frac{\left[1+(K/K_e)\Gamma_\infty\right]^{3/2}}
{\Gamma_\infty}P_0.
\end{aligned}
\label{eq:formation-input}
\end{equation}

\begin{figure*}[!t]
\centering
\includegraphics[width=\textwidth]{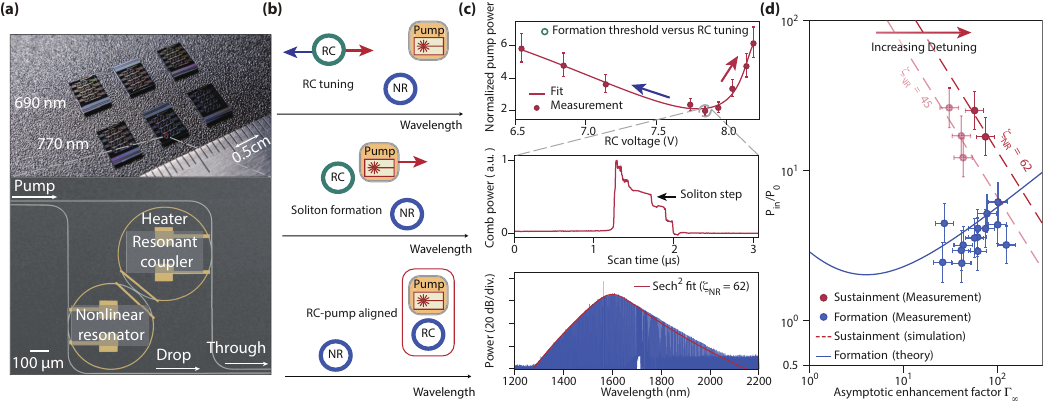}
\caption{\label{fig:experiment}
\textbf{Experimental formation and sustainment power.}
\textbf{(a)} Si$_3$N$_4$ microresonators on 690- and 770-nm-thick platforms (upper) and the coupled-resonator architecture (lower).
\textbf{(b)} Soliton-generation sequence: preset the RC resonance, scan the pump laser to form a soliton, and retune the RC toward the pump for sustainment.
\textbf{(c)} Formation power versus RC heater voltage (upper), representative soliton step (middle), and broadened soliton spectrum with a $\mathrm{sech}^2$ fit (lower).
\textbf{(d)} Measured formation and sustainment power versus $\Gamma_\infty$.}
\end{figure*}

The target soliton can be reached by continued tuning after soliton generation. Ideally, the NR detuning is increased to broaden the soliton spectrum (see Supplementary Materials), while the RC detuning is continuously adjusted to maximize the effective pump power. At large detuning, the input power required to sustain the soliton (sustainment power) should satisfy (see Supplementary Materials)
\begin{equation}
P_\mathrm{in}\geq
\frac{8\zeta_\mathrm{NR}K^2}{\pi^2K_e\mathcal G^2}P_0
=\frac{8\zeta_\mathrm{NR}}{\pi^2\Gamma_\infty}P_0.
\label{eq:sustainment-input}
\end{equation}
In this regime, increasing the RC--NR coupling lowers the sustainment power [Fig.~\ref{fig:effective-pump}(d)].

The formation and sustainment power as functions of RC--NR coupling are summarized in Fig.~\ref{fig:design}(a). For the formation power, it agrees with the numerical results obtained using full coupled-LLE model. Obviously, the minimum input power satisfying both conditions is attained at their intersection. Equating Eqs.~\eqref{eq:formation-input} and \eqref{eq:sustainment-input} therefore yields the optimal coupling,
\begin{equation}
\mathcal G_\mathrm{opt}^2
=K\left(
\frac{3\zeta_\mathrm{NR}^{2/3}}{\pi^{4/3}}-1
\right),
\label{eq:optimal-coupling}
\end{equation}
and further gives the minimum input power
\begin{equation}
P_\mathrm{in}\geq
\frac{8K\zeta_\mathrm{NR}}
{\pi^2K_e\left(3\zeta_\mathrm{NR}^{2/3}/\pi^{4/3}-1\right)}P_0
\approx
\frac{8K\zeta_\mathrm{NR}^{1/3}}
{3K_e\pi^{2/3}}P_0.
\label{eq:minimum}
\end{equation}
Note that the target soliton is assumed to operate at very large detuning ($\zeta_\mathrm{NR}\gg1$).

Using the relation $B_\mathrm{3dB}\propto\zeta_\mathrm{NR}^{1/2}$ for the 3-dB bandwidth of a soliton microcomb \cite{zhu2026power}, the large-detuning scaling in Eq.~\eqref{eq:minimum}, $P_\mathrm{in}\propto\zeta_\mathrm{NR}^{1/3}$, directly gives
\begin{equation}
P_{\mathrm{in}}\propto B_\mathrm{3dB}^{2/3}.
\label{eq:resonant-scaling}
\end{equation}
By contrast, direct pumping requires $P_\mathrm{in}^{(\mathrm{direct})}\propto\zeta_\mathrm{NR}$, corresponding to
\begin{equation}
P_\mathrm{in}^{(\mathrm{direct})}
\propto B_\mathrm{3dB}^{2}.
\label{eq:single-scaling}
\end{equation}
The advantage of resonant coupling therefore becomes increasingly pronounced as the target bandwidth increases. For example, a tenfold increase in $B_\mathrm{3dB}$ requires a hundredfold increase in input power under direct pumping, but only a factor of $10^{2/3}\simeq4.6$ with resonant coupling. This scaling is further verified numerically using the full coupled-LLE model (see Supplementary Materials). Figure~\ref{fig:design}(b) shows the minimum input power required to generate and sustain solitons as a function of their 3-dB bandwidth, in close agreement with the predicted scaling.

\textit{Experiments.---} We next test these predictions experimentally using coupled microresonators with 100-GHz-FSR NRs fabricated on 690- and 770-nm-thick Si$_3$N$_4$ platforms. Different RC--NR coupling rates are realized by varying the gap between the two resonators [Fig.~\ref{fig:experiment}(a)], while integrated heaters provide independent control of their resonance frequencies. The soliton-generation procedure, illustrated in Fig.~\ref{fig:experiment}(b), follows three steps. First, the RC resonance is set to a prescribed frequency. The pump laser is then swept from the blue to the red side of the NR resonance to initiate soliton formation. After a soliton is formed, the NR resonance is shifted to higher frequency to increase the pump--NR detuning and broaden the soliton spectrum, while the RC resonance is simultaneously tuned toward the pump frequency to increase the resonant enhancement.

\begin{figure*}[!t]
\centering
\includegraphics[width=\textwidth]{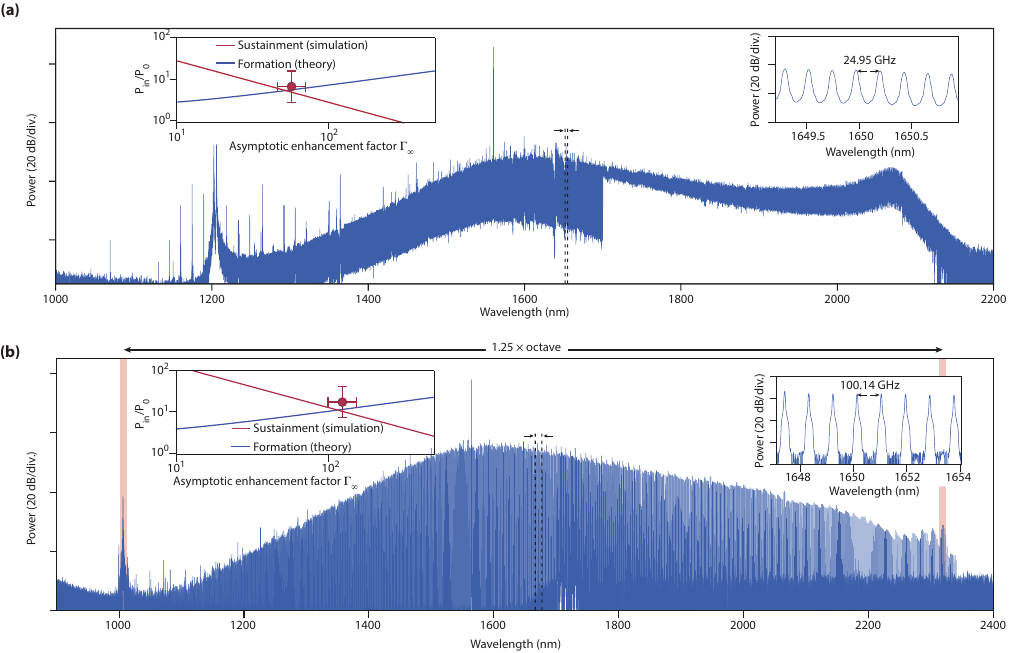}
\caption{\label{fig:devices}
\textbf{Broadband soliton microcombs at microwave repetition rates.}
\textbf{(a,b)} Measured target soliton spectra generated in devices with repetition rates of 24.95 and 100.14~GHz at on-chip pump powers of 116 and 280~mW, respectively. Left insets: formation and sustainment power versus $\Gamma_\infty$. The circles indicate the measured operating points. Right insets: magnified spectral regions resolving adjacent comb lines.}
\end{figure*}

To determine the formation power, we vary the RC heater voltage and identify the minimum input power at which characteristic soliton steps appear during the pump-laser scan [Fig.~\ref{fig:experiment}(c)]. This procedure reveals an optimal RC heater setting that minimizes the power required for soliton formation. After formation, we tune the NR to a target detuning, inferred from the measured optical spectrum, and subsequently optimize the RC resonance and input power to determine the sustainment power.

Measurements on both Si$_3$N$_4$ platforms show the same qualitative dependence of the formation power on RC--NR coupling [Fig.~\ref{fig:experiment}(d)], in agreement with the theoretical prediction. Sustainment measurements are performed on selected devices with comparable dispersion and Raman response. The light- and dark-red data correspond to the 770- and 690-nm device groups, respectively, with spectrum-inferred normalized detunings of approximately $\zeta_\mathrm{NR}=45$ and $62$. Because the Raman response can significantly modify the power required to sustain a soliton even under direct pumping, we use a generalized LLE including the measured Raman response to calculate the corresponding direct-pumping sustainment power \cite{karpov2016raman,yi2016theory}. This value serves as the baseline from which the sustainment power under resonant coupling is obtained as a function of the enhancement factor. The measured sustainment powers closely follow these simulations, confirming the opposite dependence of the two power requirements on RC--NR coupling: the formation power increases with the enhancement factor, whereas the sustainment power decreases.

\textit{Discussion.---} We finally apply the design rule to two devices operated near the predicted minimum input power. At $P_\mathrm{in}=116$ mW, the 25-GHz soliton microcomb spans approximately 1.1--2.15~\textmu m, with pronounced dispersive-wave peaks near 1.2 and 2.1~\textmu m [Fig.~\ref{fig:devices}(a)]. At $P_\mathrm{in}=280$ mW, the 100-GHz soliton microcomb spans approximately 1--2.3~\textmu m, with a strong dispersive-wave peak near 1~\textmu m [Fig.~\ref{fig:devices}(b)]. Although the spectral spans include dispersive-wave extensions and therefore differ from the 3-dB soliton bandwidth used in the scaling analysis, the 100-GHz device provide more-than-an-octave spectral coverage sufficient for direct $f$--$2f$ self-referencing \cite{diddams2020optical}. Such broadband operation is important for applications including optical clocks \cite{newman2019architecture}, optical-frequency synthesis \cite{spencer2018optical}, and low-noise microwave synthesis \cite{kudelin2024photonic,sun2024integrated,jin2025microresonator,ji2025dispersive,sun2025microcavity}.

The derived scaling law indicates only a weak power penalty for extending the soliton-microcomb bandwidth with resonant coupling. This favorable scaling suggests that the RC can be designed to target the broadest practically accessible soliton spectra. While the idealized LLE does not impose an intrinsic upper bound on the soliton bandwidth, the experimentally accessible bandwidth is ultimately constrained by effects such as the Raman self-frequency shift and higher-order dispersion \cite{wang2018stimulated}. More broadly, the effective-pump framework provides a simple description of how an auxiliary resonance modifies the excitation and dynamics of nonlinear states. Extending this framework from $\chi^{(3)}$ Kerr resonators to $\chi^{(2)}$ microresonators with coupled fundamental and second-harmonic fields \cite{xue2016second,lu2023two}, as well as to other nonlinear resonator systems, may provide a general approach for understanding and controlling nonlinear optical states through resonantly modified pump delivery.

\begin{acknowledgments}
The authors thank Xin Zhou, Jincheng Li, and Zhigang Hu for their assistance in fabrication, and Du Qian for assistance with sample photography. This work was supported by Quantum Science and Technology-National Science and
Technology Major Project (Grant No. 2021ZD0301500) and the High-performance Computing Platform of Peking University. Device fabrication was supported by the Peking University NanoOptoelectronic Fabrication Center and the Micro/nano Fabrication Laboratory of the Synergetic Extreme Condition User Facility.
\end{acknowledgments}

\noindent\textbf{Data availability.}
Data and analysis code will be deposited in Zenodo before publication and are available from the corresponding author upon reasonable request.

\bibliographystyle{apsrev4-2}
\bibliography{ref}

\begin{thebibliography}{32}%
\makeatletter
\providecommand \@ifxundefined [1]{%
 \@ifx{#1\undefined}
}%
\providecommand \@ifnum [1]{%
 \ifnum #1\expandafter \@firstoftwo
 \else \expandafter \@secondoftwo
 \fi
}%
\providecommand \@ifx [1]{%
 \ifx #1\expandafter \@firstoftwo
 \else \expandafter \@secondoftwo
 \fi
}%
\providecommand \natexlab [1]{#1}%
\providecommand \enquote  [1]{``#1''}%
\providecommand \bibnamefont  [1]{#1}%
\providecommand \bibfnamefont [1]{#1}%
\providecommand \citenamefont [1]{#1}%
\providecommand \href@noop [0]{\@secondoftwo}%
\providecommand \href [0]{\begingroup \@sanitize@url \@href}%
\providecommand \@href[1]{\@@startlink{#1}\@@href}%
\providecommand \@@href[1]{\endgroup#1\@@endlink}%
\providecommand \@sanitize@url [0]{\catcode `\\12\catcode `\$12\catcode
  `\&12\catcode `\#12\catcode `\^12\catcode `\_12\catcode `\%12\relax}%
\providecommand \@@startlink[1]{}%
\providecommand \@@endlink[0]{}%
\providecommand \url  [0]{\begingroup\@sanitize@url \@url }%
\providecommand \@url [1]{\endgroup\@href {#1}{\urlprefix }}%
\providecommand \urlprefix  [0]{URL }%
\providecommand \Eprint [0]{\href }%
\providecommand \doibase [0]{https://doi.org/}%
\providecommand \selectlanguage [0]{\@gobble}%
\providecommand \bibinfo  [0]{\@secondoftwo}%
\providecommand \bibfield  [0]{\@secondoftwo}%
\providecommand \translation [1]{[#1]}%
\providecommand \BibitemOpen [0]{}%
\providecommand \bibitemStop [0]{}%
\providecommand \bibitemNoStop [0]{.\EOS\space}%
\providecommand \EOS [0]{\spacefactor3000\relax}%
\providecommand \BibitemShut  [1]{\csname bibitem#1\endcsname}%
\let\auto@bib@innerbib\@empty
%</preamble>
\bibitem [{\citenamefont {Herr}\ \emph {et~al.}(2014)\citenamefont {Herr},
  \citenamefont {Brasch}, \citenamefont {Jost}, \citenamefont {Wang},
  \citenamefont {Kondratiev}, \citenamefont {Gorodetsky},\ and\ \citenamefont
  {Kippenberg}}]{herr2014temporal}%
  \BibitemOpen
  \bibfield  {author} {\bibinfo {author} {\bibfnamefont {T.}~\bibnamefont
  {Herr}}, \bibinfo {author} {\bibfnamefont {V.}~\bibnamefont {Brasch}},
  \bibinfo {author} {\bibfnamefont {J.}~\bibnamefont {Jost}}, \bibinfo {author}
  {\bibfnamefont {C.}~\bibnamefont {Wang}}, \bibinfo {author} {\bibfnamefont
  {N.}~\bibnamefont {Kondratiev}}, \bibinfo {author} {\bibfnamefont
  {M.}~\bibnamefont {Gorodetsky}},\ and\ \bibinfo {author} {\bibfnamefont
  {T.}~\bibnamefont {Kippenberg}},\ }\href@noop {} {\bibfield  {journal}
  {\bibinfo  {journal} {Nat. Photon.}\ }\textbf {\bibinfo {volume} {8}},\
  \bibinfo {pages} {145} (\bibinfo {year} {2014})}\BibitemShut {NoStop}%
\bibitem [{\citenamefont {Kippenberg}\ \emph {et~al.}(2018)\citenamefont
  {Kippenberg}, \citenamefont {Gaeta}, \citenamefont {Lipson},\ and\
  \citenamefont {Gorodetsky}}]{kippenberg2018dissipative}%
  \BibitemOpen
  \bibfield  {author} {\bibinfo {author} {\bibfnamefont {T.~J.}\ \bibnamefont
  {Kippenberg}}, \bibinfo {author} {\bibfnamefont {A.~L.}\ \bibnamefont
  {Gaeta}}, \bibinfo {author} {\bibfnamefont {M.}~\bibnamefont {Lipson}},\ and\
  \bibinfo {author} {\bibfnamefont {M.~L.}\ \bibnamefont {Gorodetsky}},\
  }\href@noop {} {\bibfield  {journal} {\bibinfo  {journal} {Science}\ }\textbf
  {\bibinfo {volume} {361}},\ \bibinfo {pages} {eaan8083} (\bibinfo {year}
  {2018})}\BibitemShut {NoStop}%
\bibitem [{\citenamefont {Diddams}\ \emph {et~al.}(2020)\citenamefont
  {Diddams}, \citenamefont {Vahala},\ and\ \citenamefont
  {Udem}}]{diddams2020optical}%
  \BibitemOpen
  \bibfield  {author} {\bibinfo {author} {\bibfnamefont {S.~A.}\ \bibnamefont
  {Diddams}}, \bibinfo {author} {\bibfnamefont {K.}~\bibnamefont {Vahala}},\
  and\ \bibinfo {author} {\bibfnamefont {T.}~\bibnamefont {Udem}},\ }\href@noop
  {} {\bibfield  {journal} {\bibinfo  {journal} {Science}\ }\textbf {\bibinfo
  {volume} {369}},\ \bibinfo {pages} {eaay3676} (\bibinfo {year}
  {2020})}\BibitemShut {NoStop}%
\bibitem [{\citenamefont {Del'Haye}\ \emph {et~al.}(2007)\citenamefont
  {Del'Haye}, \citenamefont {Schliesser}, \citenamefont {Arcizet},
  \citenamefont {Wilken}, \citenamefont {Holzwarth},\ and\ \citenamefont
  {Kippenberg}}]{delhaye2007optical}%
  \BibitemOpen
  \bibfield  {author} {\bibinfo {author} {\bibfnamefont {P.}~\bibnamefont
  {Del'Haye}}, \bibinfo {author} {\bibfnamefont {A.}~\bibnamefont
  {Schliesser}}, \bibinfo {author} {\bibfnamefont {O.}~\bibnamefont {Arcizet}},
  \bibinfo {author} {\bibfnamefont {T.}~\bibnamefont {Wilken}}, \bibinfo
  {author} {\bibfnamefont {R.}~\bibnamefont {Holzwarth}},\ and\ \bibinfo
  {author} {\bibfnamefont {T.}~\bibnamefont {Kippenberg}},\ }\href@noop {}
  {\bibfield  {journal} {\bibinfo  {journal} {Nature}\ }\textbf {\bibinfo
  {volume} {450}},\ \bibinfo {pages} {1214} (\bibinfo {year}
  {2007})}\BibitemShut {NoStop}%
\bibitem [{\citenamefont {Kippenberg}\ \emph {et~al.}(2011)\citenamefont
  {Kippenberg}, \citenamefont {Holzwarth},\ and\ \citenamefont
  {Diddams}}]{kippenberg2011microresonator}%
  \BibitemOpen
  \bibfield  {author} {\bibinfo {author} {\bibfnamefont {T.~J.}\ \bibnamefont
  {Kippenberg}}, \bibinfo {author} {\bibfnamefont {R.}~\bibnamefont
  {Holzwarth}},\ and\ \bibinfo {author} {\bibfnamefont {S.}~\bibnamefont
  {Diddams}},\ }\href@noop {} {\bibfield  {journal} {\bibinfo  {journal}
  {Science}\ }\textbf {\bibinfo {volume} {332}},\ \bibinfo {pages} {555}
  (\bibinfo {year} {2011})}\BibitemShut {NoStop}%
\bibitem [{\citenamefont {Suh}\ \emph {et~al.}(2016)\citenamefont {Suh},
  \citenamefont {Yang}, \citenamefont {Yang}, \citenamefont {Yi},\ and\
  \citenamefont {Vahala}}]{suh2016microresonator}%
  \BibitemOpen
  \bibfield  {author} {\bibinfo {author} {\bibfnamefont {M.-G.}\ \bibnamefont
  {Suh}}, \bibinfo {author} {\bibfnamefont {Q.-F.}\ \bibnamefont {Yang}},
  \bibinfo {author} {\bibfnamefont {K.~Y.}\ \bibnamefont {Yang}}, \bibinfo
  {author} {\bibfnamefont {X.}~\bibnamefont {Yi}},\ and\ \bibinfo {author}
  {\bibfnamefont {K.~J.}\ \bibnamefont {Vahala}},\ }\href@noop {} {\bibfield
  {journal} {\bibinfo  {journal} {Science}\ }\textbf {\bibinfo {volume}
  {354}},\ \bibinfo {pages} {600} (\bibinfo {year} {2016})}\BibitemShut
  {NoStop}%
\bibitem [{\citenamefont {Dutt}\ \emph {et~al.}(2018)\citenamefont {Dutt},
  \citenamefont {Joshi}, \citenamefont {Ji}, \citenamefont {Cardenas},
  \citenamefont {Okawachi}, \citenamefont {Luke}, \citenamefont {Gaeta},\ and\
  \citenamefont {Lipson}}]{dutt2018chip}%
  \BibitemOpen
  \bibfield  {author} {\bibinfo {author} {\bibfnamefont {A.}~\bibnamefont
  {Dutt}}, \bibinfo {author} {\bibfnamefont {C.}~\bibnamefont {Joshi}},
  \bibinfo {author} {\bibfnamefont {X.}~\bibnamefont {Ji}}, \bibinfo {author}
  {\bibfnamefont {J.}~\bibnamefont {Cardenas}}, \bibinfo {author}
  {\bibfnamefont {Y.}~\bibnamefont {Okawachi}}, \bibinfo {author}
  {\bibfnamefont {K.}~\bibnamefont {Luke}}, \bibinfo {author} {\bibfnamefont
  {A.~L.}\ \bibnamefont {Gaeta}},\ and\ \bibinfo {author} {\bibfnamefont
  {M.}~\bibnamefont {Lipson}},\ }\href@noop {} {\bibfield  {journal} {\bibinfo
  {journal} {Sci. Adv.}\ }\textbf {\bibinfo {volume} {4}},\ \bibinfo {pages}
  {e1701858} (\bibinfo {year} {2018})}\BibitemShut {NoStop}%
\bibitem [{\citenamefont {Yang}\ \emph {et~al.}(2019)\citenamefont {Yang},
  \citenamefont {Shen}, \citenamefont {Wang}, \citenamefont {Tran},
  \citenamefont {Zhang}, \citenamefont {Yang}, \citenamefont {Wu},
  \citenamefont {Bao}, \citenamefont {Bowers}, \citenamefont {Yariv},\ and\
  \citenamefont {Vahala}}]{yang2019vernier}%
  \BibitemOpen
  \bibfield  {author} {\bibinfo {author} {\bibfnamefont {Q.-F.}\ \bibnamefont
  {Yang}}, \bibinfo {author} {\bibfnamefont {B.}~\bibnamefont {Shen}}, \bibinfo
  {author} {\bibfnamefont {H.}~\bibnamefont {Wang}}, \bibinfo {author}
  {\bibfnamefont {M.}~\bibnamefont {Tran}}, \bibinfo {author} {\bibfnamefont
  {Z.}~\bibnamefont {Zhang}}, \bibinfo {author} {\bibfnamefont {K.~Y.}\
  \bibnamefont {Yang}}, \bibinfo {author} {\bibfnamefont {L.}~\bibnamefont
  {Wu}}, \bibinfo {author} {\bibfnamefont {C.}~\bibnamefont {Bao}}, \bibinfo
  {author} {\bibfnamefont {J.}~\bibnamefont {Bowers}}, \bibinfo {author}
  {\bibfnamefont {A.}~\bibnamefont {Yariv}},\ and\ \bibinfo {author}
  {\bibfnamefont {K.}~\bibnamefont {Vahala}},\ }\href@noop {} {\bibfield
  {journal} {\bibinfo  {journal} {Science}\ }\textbf {\bibinfo {volume}
  {363}},\ \bibinfo {pages} {965} (\bibinfo {year} {2019})}\BibitemShut
  {NoStop}%
\bibitem [{\citenamefont {Marin-Palomo}\ \emph {et~al.}(2017)\citenamefont
  {Marin-Palomo}, \citenamefont {Kemal}, \citenamefont {Karpov}, \citenamefont
  {Kordts}, \citenamefont {Pfeifle}, \citenamefont {Pfeiffer}, \citenamefont
  {Trocha}, \citenamefont {Wolf}, \citenamefont {Brasch}, \citenamefont
  {Anderson} \emph {et~al.}}]{marinpalomo2017microresonator}%
  \BibitemOpen
  \bibfield  {author} {\bibinfo {author} {\bibfnamefont {P.}~\bibnamefont
  {Marin-Palomo}}, \bibinfo {author} {\bibfnamefont {J.~N.}\ \bibnamefont
  {Kemal}}, \bibinfo {author} {\bibfnamefont {M.}~\bibnamefont {Karpov}},
  \bibinfo {author} {\bibfnamefont {A.}~\bibnamefont {Kordts}}, \bibinfo
  {author} {\bibfnamefont {J.}~\bibnamefont {Pfeifle}}, \bibinfo {author}
  {\bibfnamefont {M.~H.}\ \bibnamefont {Pfeiffer}}, \bibinfo {author}
  {\bibfnamefont {P.}~\bibnamefont {Trocha}}, \bibinfo {author} {\bibfnamefont
  {S.}~\bibnamefont {Wolf}}, \bibinfo {author} {\bibfnamefont {V.}~\bibnamefont
  {Brasch}}, \bibinfo {author} {\bibfnamefont {M.~H.}\ \bibnamefont
  {Anderson}}, \emph {et~al.},\ }\href@noop {} {\bibfield  {journal} {\bibinfo
  {journal} {Nature}\ }\textbf {\bibinfo {volume} {546}},\ \bibinfo {pages}
  {274} (\bibinfo {year} {2017})}\BibitemShut {NoStop}%
\bibitem [{\citenamefont {Rizzo}\ \emph {et~al.}(2022)\citenamefont {Rizzo},
  \citenamefont {Daudlin}, \citenamefont {Novick}, \citenamefont {James},
  \citenamefont {Gopal}, \citenamefont {Murthy}, \citenamefont {Cheng},
  \citenamefont {Kim}, \citenamefont {Ji}, \citenamefont {Okawachi},
  \citenamefont {Niekerk}, \citenamefont {Deenadayalan}, \citenamefont {Leake},
  \citenamefont {Fanto}, \citenamefont {Preble}, \citenamefont {Lipson},
  \citenamefont {Gaeta},\ and\ \citenamefont {Bergman}}]{rizzo2022petabit}%
  \BibitemOpen
  \bibfield  {author} {\bibinfo {author} {\bibfnamefont {A.}~\bibnamefont
  {Rizzo}}, \bibinfo {author} {\bibfnamefont {S.}~\bibnamefont {Daudlin}},
  \bibinfo {author} {\bibfnamefont {A.}~\bibnamefont {Novick}}, \bibinfo
  {author} {\bibfnamefont {A.}~\bibnamefont {James}}, \bibinfo {author}
  {\bibfnamefont {V.}~\bibnamefont {Gopal}}, \bibinfo {author} {\bibfnamefont
  {V.}~\bibnamefont {Murthy}}, \bibinfo {author} {\bibfnamefont
  {Q.}~\bibnamefont {Cheng}}, \bibinfo {author} {\bibfnamefont {B.~Y.}\
  \bibnamefont {Kim}}, \bibinfo {author} {\bibfnamefont {X.}~\bibnamefont
  {Ji}}, \bibinfo {author} {\bibfnamefont {Y.}~\bibnamefont {Okawachi}},
  \bibinfo {author} {\bibfnamefont {M.~v.}\ \bibnamefont {Niekerk}}, \bibinfo
  {author} {\bibfnamefont {V.}~\bibnamefont {Deenadayalan}}, \bibinfo {author}
  {\bibfnamefont {G.}~\bibnamefont {Leake}}, \bibinfo {author} {\bibfnamefont
  {M.~L.}\ \bibnamefont {Fanto}}, \bibinfo {author} {\bibfnamefont {S.~F.}\
  \bibnamefont {Preble}}, \bibinfo {author} {\bibfnamefont {M.}~\bibnamefont
  {Lipson}}, \bibinfo {author} {\bibfnamefont {A.~L.}\ \bibnamefont {Gaeta}},\
  and\ \bibinfo {author} {\bibfnamefont {K.}~\bibnamefont {Bergman}},\
  }\href@noop {} {\bibfield  {journal} {\bibinfo  {journal} {IEEE J. Sel.
  Topics Quantum Electron.}\ }\textbf {\bibinfo {volume} {29}},\ \bibinfo
  {pages} {1} (\bibinfo {year} {2022})}\BibitemShut {NoStop}%
\bibitem [{\citenamefont {Jørgensen}\ \emph {et~al.}(2022)\citenamefont
  {Jørgensen}, \citenamefont {Kong}, \citenamefont {Henriksen}, \citenamefont
  {Klejs}, \citenamefont {Ye}, \citenamefont {Helgason}, \citenamefont
  {Hansen}, \citenamefont {Hu}, \citenamefont {Yankov}, \citenamefont
  {Forchhammer} \emph {et~al.}}]{jrgensen2022petabit}%
  \BibitemOpen
  \bibfield  {author} {\bibinfo {author} {\bibfnamefont {A.}~\bibnamefont
  {Jørgensen}}, \bibinfo {author} {\bibfnamefont {D.}~\bibnamefont {Kong}},
  \bibinfo {author} {\bibfnamefont {M.}~\bibnamefont {Henriksen}}, \bibinfo
  {author} {\bibfnamefont {F.}~\bibnamefont {Klejs}}, \bibinfo {author}
  {\bibfnamefont {Z.}~\bibnamefont {Ye}}, \bibinfo {author} {\bibfnamefont
  {O.}~\bibnamefont {Helgason}}, \bibinfo {author} {\bibfnamefont
  {H.}~\bibnamefont {Hansen}}, \bibinfo {author} {\bibfnamefont
  {H.}~\bibnamefont {Hu}}, \bibinfo {author} {\bibfnamefont {M.}~\bibnamefont
  {Yankov}}, \bibinfo {author} {\bibfnamefont {S.}~\bibnamefont {Forchhammer}},
  \emph {et~al.},\ }\href@noop {} {\bibfield  {journal} {\bibinfo  {journal}
  {Nat. Photon.}\ }\textbf {\bibinfo {volume} {16}},\ \bibinfo {pages} {798}
  (\bibinfo {year} {2022})}\BibitemShut {NoStop}%
\bibitem [{\citenamefont {Spencer}\ \emph {et~al.}(2018)\citenamefont
  {Spencer}, \citenamefont {Drake}, \citenamefont {Briles}, \citenamefont
  {Stone}, \citenamefont {Sinclair}, \citenamefont {Fredrick}, \citenamefont
  {Li}, \citenamefont {Westly}, \citenamefont {Ilic}, \citenamefont {Bluestone}
  \emph {et~al.}}]{spencer2018optical}%
  \BibitemOpen
  \bibfield  {author} {\bibinfo {author} {\bibfnamefont {D.~T.}\ \bibnamefont
  {Spencer}}, \bibinfo {author} {\bibfnamefont {T.}~\bibnamefont {Drake}},
  \bibinfo {author} {\bibfnamefont {T.~C.}\ \bibnamefont {Briles}}, \bibinfo
  {author} {\bibfnamefont {J.}~\bibnamefont {Stone}}, \bibinfo {author}
  {\bibfnamefont {L.~C.}\ \bibnamefont {Sinclair}}, \bibinfo {author}
  {\bibfnamefont {C.}~\bibnamefont {Fredrick}}, \bibinfo {author}
  {\bibfnamefont {Q.}~\bibnamefont {Li}}, \bibinfo {author} {\bibfnamefont
  {D.}~\bibnamefont {Westly}}, \bibinfo {author} {\bibfnamefont {B.~R.}\
  \bibnamefont {Ilic}}, \bibinfo {author} {\bibfnamefont {A.}~\bibnamefont
  {Bluestone}}, \emph {et~al.},\ }\href@noop {} {\bibfield  {journal} {\bibinfo
   {journal} {Nature}\ }\textbf {\bibinfo {volume} {557}},\ \bibinfo {pages}
  {81} (\bibinfo {year} {2018})}\BibitemShut {NoStop}%
\bibitem [{\citenamefont {Newman}\ \emph {et~al.}(2019)\citenamefont {Newman},
  \citenamefont {Maurice}, \citenamefont {Drake}, \citenamefont {Stone},
  \citenamefont {Briles}, \citenamefont {Spencer}, \citenamefont {Fredrick},
  \citenamefont {Li}, \citenamefont {Westly}, \citenamefont {Ilic},
  \citenamefont {Shen}, \citenamefont {Suh}, \citenamefont {Yang},
  \citenamefont {Johnson}, \citenamefont {Johnson}, \citenamefont {Hollberg},
  \citenamefont {Vahala}, \citenamefont {Srinivasan}, \citenamefont {Diddams},
  \citenamefont {Kitching}, \citenamefont {Papp},\ and\ \citenamefont
  {Hummon}}]{newman2019architecture}%
  \BibitemOpen
  \bibfield  {author} {\bibinfo {author} {\bibfnamefont {Z.~L.}\ \bibnamefont
  {Newman}}, \bibinfo {author} {\bibfnamefont {V.}~\bibnamefont {Maurice}},
  \bibinfo {author} {\bibfnamefont {T.}~\bibnamefont {Drake}}, \bibinfo
  {author} {\bibfnamefont {J.~R.}\ \bibnamefont {Stone}}, \bibinfo {author}
  {\bibfnamefont {T.~C.}\ \bibnamefont {Briles}}, \bibinfo {author}
  {\bibfnamefont {D.~T.}\ \bibnamefont {Spencer}}, \bibinfo {author}
  {\bibfnamefont {C.}~\bibnamefont {Fredrick}}, \bibinfo {author}
  {\bibfnamefont {Q.}~\bibnamefont {Li}}, \bibinfo {author} {\bibfnamefont
  {D.}~\bibnamefont {Westly}}, \bibinfo {author} {\bibfnamefont {B.~R.}\
  \bibnamefont {Ilic}}, \bibinfo {author} {\bibfnamefont {B.}~\bibnamefont
  {Shen}}, \bibinfo {author} {\bibfnamefont {M.-G.}\ \bibnamefont {Suh}},
  \bibinfo {author} {\bibfnamefont {K.~Y.}\ \bibnamefont {Yang}}, \bibinfo
  {author} {\bibfnamefont {C.}~\bibnamefont {Johnson}}, \bibinfo {author}
  {\bibfnamefont {D.~M.~S.}\ \bibnamefont {Johnson}}, \bibinfo {author}
  {\bibfnamefont {L.}~\bibnamefont {Hollberg}}, \bibinfo {author}
  {\bibfnamefont {K.~J.}\ \bibnamefont {Vahala}}, \bibinfo {author}
  {\bibfnamefont {K.}~\bibnamefont {Srinivasan}}, \bibinfo {author}
  {\bibfnamefont {S.~A.}\ \bibnamefont {Diddams}}, \bibinfo {author}
  {\bibfnamefont {J.}~\bibnamefont {Kitching}}, \bibinfo {author}
  {\bibfnamefont {S.~B.}\ \bibnamefont {Papp}},\ and\ \bibinfo {author}
  {\bibfnamefont {M.~T.}\ \bibnamefont {Hummon}},\ }\href@noop {} {\bibfield
  {journal} {\bibinfo  {journal} {Optica}\ }\textbf {\bibinfo {volume} {6}},\
  \bibinfo {pages} {680} (\bibinfo {year} {2019})}\BibitemShut {NoStop}%
\bibitem [{\citenamefont {Lucas}\ \emph {et~al.}(2017)\citenamefont {Lucas},
  \citenamefont {Guo}, \citenamefont {Jost}, \citenamefont {Karpov},\ and\
  \citenamefont {Kippenberg}}]{lucas2017detuning}%
  \BibitemOpen
  \bibfield  {author} {\bibinfo {author} {\bibfnamefont {E.}~\bibnamefont
  {Lucas}}, \bibinfo {author} {\bibfnamefont {H.}~\bibnamefont {Guo}}, \bibinfo
  {author} {\bibfnamefont {J.~D.}\ \bibnamefont {Jost}}, \bibinfo {author}
  {\bibfnamefont {M.}~\bibnamefont {Karpov}},\ and\ \bibinfo {author}
  {\bibfnamefont {T.~J.}\ \bibnamefont {Kippenberg}},\ }\href@noop {}
  {\bibfield  {journal} {\bibinfo  {journal} {Phys. Rev. A}\ }\textbf {\bibinfo
  {volume} {95}},\ \bibinfo {pages} {043822} (\bibinfo {year}
  {2017})}\BibitemShut {NoStop}%
\bibitem [{\citenamefont {Li}\ \emph {et~al.}(2018)\citenamefont {Li},
  \citenamefont {Shen}, \citenamefont {Wang}, \citenamefont {Yang},
  \citenamefont {Yi}, \citenamefont {Yang}, \citenamefont {Zhou},\ and\
  \citenamefont {Vahala}}]{li2018universal}%
  \BibitemOpen
  \bibfield  {author} {\bibinfo {author} {\bibfnamefont {X.}~\bibnamefont
  {Li}}, \bibinfo {author} {\bibfnamefont {B.}~\bibnamefont {Shen}}, \bibinfo
  {author} {\bibfnamefont {H.}~\bibnamefont {Wang}}, \bibinfo {author}
  {\bibfnamefont {K.~Y.}\ \bibnamefont {Yang}}, \bibinfo {author}
  {\bibfnamefont {X.}~\bibnamefont {Yi}}, \bibinfo {author} {\bibfnamefont
  {Q.-F.}\ \bibnamefont {Yang}}, \bibinfo {author} {\bibfnamefont
  {Z.}~\bibnamefont {Zhou}},\ and\ \bibinfo {author} {\bibfnamefont
  {K.}~\bibnamefont {Vahala}},\ }\href@noop {} {\bibfield  {journal} {\bibinfo
  {journal} {Opt. Lett.}\ }\textbf {\bibinfo {volume} {43}},\ \bibinfo {pages}
  {2567} (\bibinfo {year} {2018})}\BibitemShut {NoStop}%
\bibitem [{\citenamefont {Stern}\ \emph {et~al.}(2018)\citenamefont {Stern},
  \citenamefont {Ji}, \citenamefont {Okawachi}, \citenamefont {Gaeta},\ and\
  \citenamefont {Lipson}}]{stern2018battery}%
  \BibitemOpen
  \bibfield  {author} {\bibinfo {author} {\bibfnamefont {B.}~\bibnamefont
  {Stern}}, \bibinfo {author} {\bibfnamefont {X.}~\bibnamefont {Ji}}, \bibinfo
  {author} {\bibfnamefont {Y.}~\bibnamefont {Okawachi}}, \bibinfo {author}
  {\bibfnamefont {A.~L.}\ \bibnamefont {Gaeta}},\ and\ \bibinfo {author}
  {\bibfnamefont {M.}~\bibnamefont {Lipson}},\ }\href@noop {} {\bibfield
  {journal} {\bibinfo  {journal} {Nature}\ }\textbf {\bibinfo {volume} {562}},\
  \bibinfo {pages} {401} (\bibinfo {year} {2018})}\BibitemShut {NoStop}%
\bibitem [{\citenamefont {Raja}\ \emph {et~al.}(2019)\citenamefont {Raja},
  \citenamefont {Voloshin}, \citenamefont {Guo}, \citenamefont {Agafonova},
  \citenamefont {Liu}, \citenamefont {Gorodnitskiy}, \citenamefont {Karpov},
  \citenamefont {Pavlov}, \citenamefont {Lucas}, \citenamefont {Galiev} \emph
  {et~al.}}]{raja2019electrically}%
  \BibitemOpen
  \bibfield  {author} {\bibinfo {author} {\bibfnamefont {A.~S.}\ \bibnamefont
  {Raja}}, \bibinfo {author} {\bibfnamefont {A.~S.}\ \bibnamefont {Voloshin}},
  \bibinfo {author} {\bibfnamefont {H.}~\bibnamefont {Guo}}, \bibinfo {author}
  {\bibfnamefont {S.~E.}\ \bibnamefont {Agafonova}}, \bibinfo {author}
  {\bibfnamefont {J.}~\bibnamefont {Liu}}, \bibinfo {author} {\bibfnamefont
  {A.~S.}\ \bibnamefont {Gorodnitskiy}}, \bibinfo {author} {\bibfnamefont
  {M.}~\bibnamefont {Karpov}}, \bibinfo {author} {\bibfnamefont {N.~G.}\
  \bibnamefont {Pavlov}}, \bibinfo {author} {\bibfnamefont {E.}~\bibnamefont
  {Lucas}}, \bibinfo {author} {\bibfnamefont {R.~R.}\ \bibnamefont {Galiev}},
  \emph {et~al.},\ }\href@noop {} {\bibfield  {journal} {\bibinfo  {journal}
  {Nat. Commun.}\ }\textbf {\bibinfo {volume} {10}},\ \bibinfo {pages} {680}
  (\bibinfo {year} {2019})}\BibitemShut {NoStop}%
\bibitem [{\citenamefont {Shen}\ \emph {et~al.}(2020)\citenamefont {Shen},
  \citenamefont {Chang}, \citenamefont {Liu}, \citenamefont {Wang},
  \citenamefont {Yang}, \citenamefont {Xiang}, \citenamefont {Wang},
  \citenamefont {He}, \citenamefont {Liu}, \citenamefont {Xie}, \citenamefont
  {Guo}, \citenamefont {Kinghorn}, \citenamefont {Wu}, \citenamefont {Ji},
  \citenamefont {Kippenberg}, \citenamefont {Vahala},\ and\ \citenamefont
  {Bowers}}]{shen2020integrated}%
  \BibitemOpen
  \bibfield  {author} {\bibinfo {author} {\bibfnamefont {B.}~\bibnamefont
  {Shen}}, \bibinfo {author} {\bibfnamefont {L.}~\bibnamefont {Chang}},
  \bibinfo {author} {\bibfnamefont {J.}~\bibnamefont {Liu}}, \bibinfo {author}
  {\bibfnamefont {H.}~\bibnamefont {Wang}}, \bibinfo {author} {\bibfnamefont
  {Q.-F.}\ \bibnamefont {Yang}}, \bibinfo {author} {\bibfnamefont
  {C.}~\bibnamefont {Xiang}}, \bibinfo {author} {\bibfnamefont {R.~N.}\
  \bibnamefont {Wang}}, \bibinfo {author} {\bibfnamefont {J.}~\bibnamefont
  {He}}, \bibinfo {author} {\bibfnamefont {T.}~\bibnamefont {Liu}}, \bibinfo
  {author} {\bibfnamefont {W.}~\bibnamefont {Xie}}, \bibinfo {author}
  {\bibfnamefont {J.}~\bibnamefont {Guo}}, \bibinfo {author} {\bibfnamefont
  {D.}~\bibnamefont {Kinghorn}}, \bibinfo {author} {\bibfnamefont
  {L.}~\bibnamefont {Wu}}, \bibinfo {author} {\bibfnamefont {Q.-X.}\
  \bibnamefont {Ji}}, \bibinfo {author} {\bibfnamefont {T.~J.}\ \bibnamefont
  {Kippenberg}}, \bibinfo {author} {\bibfnamefont {K.}~\bibnamefont {Vahala}},\
  and\ \bibinfo {author} {\bibfnamefont {J.~E.}\ \bibnamefont {Bowers}},\
  }\href@noop {} {\bibfield  {journal} {\bibinfo  {journal} {Nature}\ }\textbf
  {\bibinfo {volume} {582}},\ \bibinfo {pages} {365} (\bibinfo {year}
  {2020})}\BibitemShut {NoStop}%
\bibitem [{\citenamefont {Xiang}\ \emph {et~al.}(2021)\citenamefont {Xiang},
  \citenamefont {Liu}, \citenamefont {Guo}, \citenamefont {Chang},
  \citenamefont {Wang}, \citenamefont {Weng}, \citenamefont {Peters},
  \citenamefont {Xie}, \citenamefont {Zhang}, \citenamefont {Riemensberger}
  \emph {et~al.}}]{xiang2021laser}%
  \BibitemOpen
  \bibfield  {author} {\bibinfo {author} {\bibfnamefont {C.}~\bibnamefont
  {Xiang}}, \bibinfo {author} {\bibfnamefont {J.}~\bibnamefont {Liu}}, \bibinfo
  {author} {\bibfnamefont {J.}~\bibnamefont {Guo}}, \bibinfo {author}
  {\bibfnamefont {L.}~\bibnamefont {Chang}}, \bibinfo {author} {\bibfnamefont
  {R.~N.}\ \bibnamefont {Wang}}, \bibinfo {author} {\bibfnamefont
  {W.}~\bibnamefont {Weng}}, \bibinfo {author} {\bibfnamefont {J.}~\bibnamefont
  {Peters}}, \bibinfo {author} {\bibfnamefont {W.}~\bibnamefont {Xie}},
  \bibinfo {author} {\bibfnamefont {Z.}~\bibnamefont {Zhang}}, \bibinfo
  {author} {\bibfnamefont {J.}~\bibnamefont {Riemensberger}}, \emph {et~al.},\
  }\href@noop {} {\bibfield  {journal} {\bibinfo  {journal} {Science}\ }\textbf
  {\bibinfo {volume} {373}},\ \bibinfo {pages} {99} (\bibinfo {year}
  {2021})}\BibitemShut {NoStop}%
\bibitem [{\citenamefont {Yang}\ \emph {et~al.}(2024)\citenamefont {Yang},
  \citenamefont {Hu}, \citenamefont {Torres-Company},\ and\ \citenamefont
  {Vahala}}]{yang2024efficient}%
  \BibitemOpen
  \bibfield  {author} {\bibinfo {author} {\bibfnamefont {Q.-F.}\ \bibnamefont
  {Yang}}, \bibinfo {author} {\bibfnamefont {Y.}~\bibnamefont {Hu}}, \bibinfo
  {author} {\bibfnamefont {V.}~\bibnamefont {Torres-Company}},\ and\ \bibinfo
  {author} {\bibfnamefont {K.}~\bibnamefont {Vahala}},\ }\href@noop {}
  {\bibfield  {journal} {\bibinfo  {journal} {eLight}\ }\textbf {\bibinfo
  {volume} {4}},\ \bibinfo {pages} {18} (\bibinfo {year} {2024})}\BibitemShut
  {NoStop}%
\bibitem [{\citenamefont {Xue}\ \emph {et~al.}(2019)\citenamefont {Xue},
  \citenamefont {Zheng},\ and\ \citenamefont {Zhou}}]{xue2019super}%
  \BibitemOpen
  \bibfield  {author} {\bibinfo {author} {\bibfnamefont {X.}~\bibnamefont
  {Xue}}, \bibinfo {author} {\bibfnamefont {X.}~\bibnamefont {Zheng}},\ and\
  \bibinfo {author} {\bibfnamefont {B.}~\bibnamefont {Zhou}},\ }\href@noop {}
  {\bibfield  {journal} {\bibinfo  {journal} {Nat. Photon.}\ }\textbf {\bibinfo
  {volume} {13}},\ \bibinfo {pages} {616} (\bibinfo {year} {2019})}\BibitemShut
  {NoStop}%
\bibitem [{\citenamefont {Zhu}\ \emph {et~al.}(2026)\citenamefont {Zhu},
  \citenamefont {Luo}, \citenamefont {Wang}, \citenamefont {Wang},
  \citenamefont {Xu}, \citenamefont {Qian}, \citenamefont {Cheng},
  \citenamefont {Wang}, \citenamefont {Luo}, \citenamefont {Liu}, \citenamefont
  {Jin}, \citenamefont {Xie}, \citenamefont {Zhou}, \citenamefont {Wang},
  \citenamefont {Liu}, \citenamefont {Cao}, \citenamefont {Wang}, \citenamefont
  {Tang}, \citenamefont {Gong}, \citenamefont {Li},\ and\ \citenamefont
  {Yang}}]{zhu2026power}%
  \BibitemOpen
  \bibfield  {author} {\bibinfo {author} {\bibfnamefont {K.}~\bibnamefont
  {Zhu}}, \bibinfo {author} {\bibfnamefont {X.}~\bibnamefont {Luo}}, \bibinfo
  {author} {\bibfnamefont {Y.}~\bibnamefont {Wang}}, \bibinfo {author}
  {\bibfnamefont {Z.}~\bibnamefont {Wang}}, \bibinfo {author} {\bibfnamefont
  {T.}~\bibnamefont {Xu}}, \bibinfo {author} {\bibfnamefont {D.}~\bibnamefont
  {Qian}}, \bibinfo {author} {\bibfnamefont {Y.}~\bibnamefont {Cheng}},
  \bibinfo {author} {\bibfnamefont {J.}~\bibnamefont {Wang}}, \bibinfo {author}
  {\bibfnamefont {H.}~\bibnamefont {Luo}}, \bibinfo {author} {\bibfnamefont
  {Y.}~\bibnamefont {Liu}}, \bibinfo {author} {\bibfnamefont {X.}~\bibnamefont
  {Jin}}, \bibinfo {author} {\bibfnamefont {Z.}~\bibnamefont {Xie}}, \bibinfo
  {author} {\bibfnamefont {X.}~\bibnamefont {Zhou}}, \bibinfo {author}
  {\bibfnamefont {M.}~\bibnamefont {Wang}}, \bibinfo {author} {\bibfnamefont
  {J.-F.}\ \bibnamefont {Liu}}, \bibinfo {author} {\bibfnamefont
  {X.}~\bibnamefont {Cao}}, \bibinfo {author} {\bibfnamefont {T.}~\bibnamefont
  {Wang}}, \bibinfo {author} {\bibfnamefont {S.-J.}\ \bibnamefont {Tang}},
  \bibinfo {author} {\bibfnamefont {Q.}~\bibnamefont {Gong}}, \bibinfo {author}
  {\bibfnamefont {B.-B.}\ \bibnamefont {Li}},\ and\ \bibinfo {author}
  {\bibfnamefont {Q.-F.}\ \bibnamefont {Yang}},\ }\href@noop {} {\bibfield
  {journal} {\bibinfo  {journal} {Light: Sci. Appl.}\ }\textbf {\bibinfo
  {volume} {15}},\ \bibinfo {pages} {185} (\bibinfo {year} {2026})}\BibitemShut
  {NoStop}%
\bibitem [{\citenamefont {Karpov}\ \emph {et~al.}(2016)\citenamefont {Karpov},
  \citenamefont {Guo}, \citenamefont {Kordts}, \citenamefont {Brasch},
  \citenamefont {Pfeiffer}, \citenamefont {Zervas}, \citenamefont
  {Geiselmann},\ and\ \citenamefont {Kippenberg}}]{karpov2016raman}%
  \BibitemOpen
  \bibfield  {author} {\bibinfo {author} {\bibfnamefont {M.}~\bibnamefont
  {Karpov}}, \bibinfo {author} {\bibfnamefont {H.}~\bibnamefont {Guo}},
  \bibinfo {author} {\bibfnamefont {A.}~\bibnamefont {Kordts}}, \bibinfo
  {author} {\bibfnamefont {V.}~\bibnamefont {Brasch}}, \bibinfo {author}
  {\bibfnamefont {M.~H.}\ \bibnamefont {Pfeiffer}}, \bibinfo {author}
  {\bibfnamefont {M.}~\bibnamefont {Zervas}}, \bibinfo {author} {\bibfnamefont
  {M.}~\bibnamefont {Geiselmann}},\ and\ \bibinfo {author} {\bibfnamefont
  {T.~J.}\ \bibnamefont {Kippenberg}},\ }\href@noop {} {\bibfield  {journal}
  {\bibinfo  {journal} {Phys. Rev. Lett.}\ }\textbf {\bibinfo {volume} {116}},\
  \bibinfo {pages} {103902} (\bibinfo {year} {2016})}\BibitemShut {NoStop}%
\bibitem [{\citenamefont {Yi}\ \emph {et~al.}(2016)\citenamefont {Yi},
  \citenamefont {Yang}, \citenamefont {Yang},\ and\ \citenamefont
  {Vahala}}]{yi2016theory}%
  \BibitemOpen
  \bibfield  {author} {\bibinfo {author} {\bibfnamefont {X.}~\bibnamefont
  {Yi}}, \bibinfo {author} {\bibfnamefont {Q.-F.}\ \bibnamefont {Yang}},
  \bibinfo {author} {\bibfnamefont {K.~Y.}\ \bibnamefont {Yang}},\ and\
  \bibinfo {author} {\bibfnamefont {K.}~\bibnamefont {Vahala}},\ }\href@noop {}
  {\bibfield  {journal} {\bibinfo  {journal} {Opt. Lett.}\ }\textbf {\bibinfo
  {volume} {41}},\ \bibinfo {pages} {3419} (\bibinfo {year}
  {2016})}\BibitemShut {NoStop}%
\bibitem [{\citenamefont {Kudelin}\ \emph {et~al.}(2024)\citenamefont
  {Kudelin}, \citenamefont {Groman}, \citenamefont {Ji}, \citenamefont {Guo},
  \citenamefont {Kelleher}, \citenamefont {Lee}, \citenamefont {Nakamura},
  \citenamefont {McLemore}, \citenamefont {Shirmohammadi}, \citenamefont
  {Hanifi} \emph {et~al.}}]{kudelin2024photonic}%
  \BibitemOpen
  \bibfield  {author} {\bibinfo {author} {\bibfnamefont {I.}~\bibnamefont
  {Kudelin}}, \bibinfo {author} {\bibfnamefont {W.}~\bibnamefont {Groman}},
  \bibinfo {author} {\bibfnamefont {Q.-X.}\ \bibnamefont {Ji}}, \bibinfo
  {author} {\bibfnamefont {J.}~\bibnamefont {Guo}}, \bibinfo {author}
  {\bibfnamefont {M.~L.}\ \bibnamefont {Kelleher}}, \bibinfo {author}
  {\bibfnamefont {D.}~\bibnamefont {Lee}}, \bibinfo {author} {\bibfnamefont
  {T.}~\bibnamefont {Nakamura}}, \bibinfo {author} {\bibfnamefont {C.~A.}\
  \bibnamefont {McLemore}}, \bibinfo {author} {\bibfnamefont {P.}~\bibnamefont
  {Shirmohammadi}}, \bibinfo {author} {\bibfnamefont {S.}~\bibnamefont
  {Hanifi}}, \emph {et~al.},\ }\href@noop {} {\bibfield  {journal} {\bibinfo
  {journal} {Nature}\ }\textbf {\bibinfo {volume} {627}},\ \bibinfo {pages}
  {534} (\bibinfo {year} {2024})}\BibitemShut {NoStop}%
\bibitem [{\citenamefont {Sun}\ \emph {et~al.}(2024)\citenamefont {Sun},
  \citenamefont {Wang}, \citenamefont {Liu}, \citenamefont {Harrington},
  \citenamefont {Tabatabaei}, \citenamefont {Liu}, \citenamefont {Wang},
  \citenamefont {Hanifi}, \citenamefont {Morgan}, \citenamefont {Jahanbozorgi}
  \emph {et~al.}}]{sun2024integrated}%
  \BibitemOpen
  \bibfield  {author} {\bibinfo {author} {\bibfnamefont {S.}~\bibnamefont
  {Sun}}, \bibinfo {author} {\bibfnamefont {B.}~\bibnamefont {Wang}}, \bibinfo
  {author} {\bibfnamefont {K.}~\bibnamefont {Liu}}, \bibinfo {author}
  {\bibfnamefont {M.~W.}\ \bibnamefont {Harrington}}, \bibinfo {author}
  {\bibfnamefont {F.}~\bibnamefont {Tabatabaei}}, \bibinfo {author}
  {\bibfnamefont {R.}~\bibnamefont {Liu}}, \bibinfo {author} {\bibfnamefont
  {J.}~\bibnamefont {Wang}}, \bibinfo {author} {\bibfnamefont {S.}~\bibnamefont
  {Hanifi}}, \bibinfo {author} {\bibfnamefont {J.~S.}\ \bibnamefont {Morgan}},
  \bibinfo {author} {\bibfnamefont {M.}~\bibnamefont {Jahanbozorgi}}, \emph
  {et~al.},\ }\href@noop {} {\bibfield  {journal} {\bibinfo  {journal}
  {Nature}\ }\textbf {\bibinfo {volume} {627}},\ \bibinfo {pages} {540}
  (\bibinfo {year} {2024})}\BibitemShut {NoStop}%
\bibitem [{\citenamefont {Jin}\ \emph {et~al.}(2025)\citenamefont {Jin},
  \citenamefont {Xie}, \citenamefont {Zhang}, \citenamefont {Hou},
  \citenamefont {Wu}, \citenamefont {Zhang}, \citenamefont {Zhang},
  \citenamefont {Chang}, \citenamefont {Gong},\ and\ \citenamefont
  {Yang}}]{jin2025microresonator}%
  \BibitemOpen
  \bibfield  {author} {\bibinfo {author} {\bibfnamefont {X.}~\bibnamefont
  {Jin}}, \bibinfo {author} {\bibfnamefont {Z.}~\bibnamefont {Xie}}, \bibinfo
  {author} {\bibfnamefont {X.}~\bibnamefont {Zhang}}, \bibinfo {author}
  {\bibfnamefont {H.}~\bibnamefont {Hou}}, \bibinfo {author} {\bibfnamefont
  {B.}~\bibnamefont {Wu}}, \bibinfo {author} {\bibfnamefont {F.}~\bibnamefont
  {Zhang}}, \bibinfo {author} {\bibfnamefont {X.}~\bibnamefont {Zhang}},
  \bibinfo {author} {\bibfnamefont {L.}~\bibnamefont {Chang}}, \bibinfo
  {author} {\bibfnamefont {Q.}~\bibnamefont {Gong}},\ and\ \bibinfo {author}
  {\bibfnamefont {Q.-F.}\ \bibnamefont {Yang}},\ }\href@noop {} {\bibfield
  {journal} {\bibinfo  {journal} {Nat. Photon.}\ }\textbf {\bibinfo {volume}
  {19}},\ \bibinfo {pages} {630} (\bibinfo {year} {2025})}\BibitemShut
  {NoStop}%
\bibitem [{\citenamefont {Ji}\ \emph {et~al.}(2025)\citenamefont {Ji},
  \citenamefont {Zhang}, \citenamefont {Savchenkov}, \citenamefont {Liu},
  \citenamefont {Sun}, \citenamefont {Jin}, \citenamefont {Guo}, \citenamefont
  {Peters}, \citenamefont {Wu}, \citenamefont {Feshali}, \citenamefont
  {Paniccia}, \citenamefont {Ilchenko}, \citenamefont {Bowers}, \citenamefont
  {Matsko},\ and\ \citenamefont {Vahala}}]{ji2025dispersive}%
  \BibitemOpen
  \bibfield  {author} {\bibinfo {author} {\bibfnamefont {Q.-X.}\ \bibnamefont
  {Ji}}, \bibinfo {author} {\bibfnamefont {W.}~\bibnamefont {Zhang}}, \bibinfo
  {author} {\bibfnamefont {A.}~\bibnamefont {Savchenkov}}, \bibinfo {author}
  {\bibfnamefont {P.}~\bibnamefont {Liu}}, \bibinfo {author} {\bibfnamefont
  {S.}~\bibnamefont {Sun}}, \bibinfo {author} {\bibfnamefont {W.}~\bibnamefont
  {Jin}}, \bibinfo {author} {\bibfnamefont {J.}~\bibnamefont {Guo}}, \bibinfo
  {author} {\bibfnamefont {J.}~\bibnamefont {Peters}}, \bibinfo {author}
  {\bibfnamefont {L.}~\bibnamefont {Wu}}, \bibinfo {author} {\bibfnamefont
  {A.}~\bibnamefont {Feshali}}, \bibinfo {author} {\bibfnamefont
  {M.}~\bibnamefont {Paniccia}}, \bibinfo {author} {\bibfnamefont
  {V.}~\bibnamefont {Ilchenko}}, \bibinfo {author} {\bibfnamefont
  {J.}~\bibnamefont {Bowers}}, \bibinfo {author} {\bibfnamefont
  {A.}~\bibnamefont {Matsko}},\ and\ \bibinfo {author} {\bibfnamefont
  {K.}~\bibnamefont {Vahala}},\ }\href@noop {} {\bibfield  {journal} {\bibinfo
  {journal} {Nat. Photon.}\ }\textbf {\bibinfo {volume} {19}},\ \bibinfo
  {pages} {624} (\bibinfo {year} {2025})}\BibitemShut {NoStop}%
\bibitem [{\citenamefont {Sun}\ \emph {et~al.}(2025)\citenamefont {Sun},
  \citenamefont {Harrington}, \citenamefont {Tabatabaei}, \citenamefont
  {Hanifi}, \citenamefont {Liu}, \citenamefont {Wang}, \citenamefont {Wang},
  \citenamefont {Yang}, \citenamefont {Liu}, \citenamefont {Morgan},
  \citenamefont {Bowers}, \citenamefont {Morton}, \citenamefont {Nelson},
  \citenamefont {Beling}, \citenamefont {Blumenthal},\ and\ \citenamefont
  {Yi}}]{sun2025microcavity}%
  \BibitemOpen
  \bibfield  {author} {\bibinfo {author} {\bibfnamefont {S.}~\bibnamefont
  {Sun}}, \bibinfo {author} {\bibfnamefont {M.}~\bibnamefont {Harrington}},
  \bibinfo {author} {\bibfnamefont {F.}~\bibnamefont {Tabatabaei}}, \bibinfo
  {author} {\bibfnamefont {S.}~\bibnamefont {Hanifi}}, \bibinfo {author}
  {\bibfnamefont {K.}~\bibnamefont {Liu}}, \bibinfo {author} {\bibfnamefont
  {J.}~\bibnamefont {Wang}}, \bibinfo {author} {\bibfnamefont {B.}~\bibnamefont
  {Wang}}, \bibinfo {author} {\bibfnamefont {Z.}~\bibnamefont {Yang}}, \bibinfo
  {author} {\bibfnamefont {R.}~\bibnamefont {Liu}}, \bibinfo {author}
  {\bibfnamefont {J.}~\bibnamefont {Morgan}}, \bibinfo {author} {\bibfnamefont
  {S.~M.}\ \bibnamefont {Bowers}}, \bibinfo {author} {\bibfnamefont {P.~A.}\
  \bibnamefont {Morton}}, \bibinfo {author} {\bibfnamefont {K.~D.}\
  \bibnamefont {Nelson}}, \bibinfo {author} {\bibfnamefont {A.}~\bibnamefont
  {Beling}}, \bibinfo {author} {\bibfnamefont {D.~J.}\ \bibnamefont
  {Blumenthal}},\ and\ \bibinfo {author} {\bibfnamefont {X.}~\bibnamefont
  {Yi}},\ }\href@noop {} {\bibfield  {journal} {\bibinfo  {journal} {Nat.
  Photon.}\ }\textbf {\bibinfo {volume} {19}},\ \bibinfo {pages} {637}
  (\bibinfo {year} {2025})}\BibitemShut {NoStop}%
\bibitem [{\citenamefont {Wang}\ \emph {et~al.}(2018)\citenamefont {Wang},
  \citenamefont {Anderson}, \citenamefont {Coen}, \citenamefont {Murdoch},\
  and\ \citenamefont {Erkintalo}}]{wang2018stimulated}%
  \BibitemOpen
  \bibfield  {author} {\bibinfo {author} {\bibfnamefont {Y.}~\bibnamefont
  {Wang}}, \bibinfo {author} {\bibfnamefont {M.}~\bibnamefont {Anderson}},
  \bibinfo {author} {\bibfnamefont {S.}~\bibnamefont {Coen}}, \bibinfo {author}
  {\bibfnamefont {S.~G.}\ \bibnamefont {Murdoch}},\ and\ \bibinfo {author}
  {\bibfnamefont {M.}~\bibnamefont {Erkintalo}},\ }\href@noop {} {\bibfield
  {journal} {\bibinfo  {journal} {Phys. Rev. Lett.}\ }\textbf {\bibinfo
  {volume} {120}},\ \bibinfo {pages} {053902} (\bibinfo {year}
  {2018})}\BibitemShut {NoStop}%
\bibitem [{\citenamefont {Xue}\ \emph {et~al.}(2016)\citenamefont {Xue},
  \citenamefont {Leo}, \citenamefont {Xuan}, \citenamefont
  {Jaramillo-Villegas}, \citenamefont {Wang}, \citenamefont {Leaird},
  \citenamefont {Erkintalo}, \citenamefont {Qi},\ and\ \citenamefont
  {Weiner}}]{xue2016second}%
  \BibitemOpen
  \bibfield  {author} {\bibinfo {author} {\bibfnamefont {X.}~\bibnamefont
  {Xue}}, \bibinfo {author} {\bibfnamefont {F.}~\bibnamefont {Leo}}, \bibinfo
  {author} {\bibfnamefont {Y.}~\bibnamefont {Xuan}}, \bibinfo {author}
  {\bibfnamefont {J.~A.}\ \bibnamefont {Jaramillo-Villegas}}, \bibinfo {author}
  {\bibfnamefont {P.-H.}\ \bibnamefont {Wang}}, \bibinfo {author}
  {\bibfnamefont {D.~E.}\ \bibnamefont {Leaird}}, \bibinfo {author}
  {\bibfnamefont {M.}~\bibnamefont {Erkintalo}}, \bibinfo {author}
  {\bibfnamefont {M.}~\bibnamefont {Qi}},\ and\ \bibinfo {author}
  {\bibfnamefont {A.~M.}\ \bibnamefont {Weiner}},\ }\href@noop {} {\bibfield
  {journal} {\bibinfo  {journal} {Light: Sci. Appl.}\ }\textbf {\bibinfo
  {volume} {6}},\ \bibinfo {pages} {e16253} (\bibinfo {year}
  {2016})}\BibitemShut {NoStop}%
\bibitem [{\citenamefont {Lu}\ \emph {et~al.}(2023)\citenamefont {Lu},
  \citenamefont {Puzyrev}, \citenamefont {Pankratov}, \citenamefont {Skryabin},
  \citenamefont {Yang}, \citenamefont {Gong}, \citenamefont {Surya},\ and\
  \citenamefont {Tang}}]{lu2023two}%
  \BibitemOpen
  \bibfield  {author} {\bibinfo {author} {\bibfnamefont {J.}~\bibnamefont
  {Lu}}, \bibinfo {author} {\bibfnamefont {D.~N.}\ \bibnamefont {Puzyrev}},
  \bibinfo {author} {\bibfnamefont {V.~V.}\ \bibnamefont {Pankratov}}, \bibinfo
  {author} {\bibfnamefont {D.~V.}\ \bibnamefont {Skryabin}}, \bibinfo {author}
  {\bibfnamefont {F.}~\bibnamefont {Yang}}, \bibinfo {author} {\bibfnamefont
  {Z.}~\bibnamefont {Gong}}, \bibinfo {author} {\bibfnamefont {J.~B.}\
  \bibnamefont {Surya}},\ and\ \bibinfo {author} {\bibfnamefont {H.~X.}\
  \bibnamefont {Tang}},\ }\href@noop {} {\bibfield  {journal} {\bibinfo
  {journal} {Nat. Commun.}\ }\textbf {\bibinfo {volume} {14}},\ \bibinfo
  {pages} {2798} (\bibinfo {year} {2023})}\BibitemShut {NoStop}%
\end{thebibliography}%


\begin{thebibliography}{10}
\expandafter\ifx\csname url\endcsname\relax
  \def\url#1{\texttt{#1}}\fi
\expandafter\ifx\csname urlprefix\endcsname\relax\def\urlprefix{URL }\fi
\providecommand{\bibinfo}[2]{#2}
\providecommand{\eprint}[2][]{\url{#2}}

\bibitem{JSUVFVVT}
\bibinfo{author}{Zhu, K.} \emph{et~al.}
\newblock \bibinfo{title}{Power-efficient ultra-broadband soliton microcombs in
  resonantly-coupled microresonators}.
\newblock \emph{\bibinfo{journal}{Light Sci. Appl.}}
  \textbf{\bibinfo{volume}{15}}, \bibinfo{pages}{185} (\bibinfo{year}{2026}).

\bibitem{RNZK9EYV}
\bibinfo{author}{Haus, H.~A.} \& \bibinfo{author}{Huang, W.~P.}
\newblock \bibinfo{title}{Coupled-mode theory}.
\newblock \emph{\bibinfo{journal}{Proc. IEEE}} \textbf{\bibinfo{volume}{79}},
  \bibinfo{pages}{1505--1518} (\bibinfo{year}{1991}).

\bibitem{KX4LXUPF}
\bibinfo{author}{Chembo, Y.~K.} \& \bibinfo{author}{Menyuk, C.~R.}
\newblock \bibinfo{title}{Spatiotemporal lugiato-lefever formalism for
  kerr-comb generation in whispering-gallery-mode resonators}.
\newblock \emph{\bibinfo{journal}{Phys. Rev. A}} \textbf{\bibinfo{volume}{87}},
  \bibinfo{pages}{053852} (\bibinfo{year}{2013}).

\bibitem{76NN4HTK}
\bibinfo{author}{Godey, C.}, \bibinfo{author}{Balakireva, I.~V.},
  \bibinfo{author}{Coillet, A.} \& \bibinfo{author}{Chembo, Y.~K.}
\newblock \bibinfo{title}{Stability analysis of the spatiotemporal
  {L}ugiato-{L}efever model for {K}err optical frequency combs in the anomalous
  and normal dispersion regimes}.
\newblock \emph{\bibinfo{journal}{Phys. Rev. A}} \textbf{\bibinfo{volume}{89}},
  \bibinfo{pages}{063814} (\bibinfo{year}{2014}).

\bibitem{GQ4W33CB}
\bibinfo{author}{Herr, T.} \emph{et~al.}
\newblock \bibinfo{title}{Temporal solitons in optical microresonators}.
\newblock \emph{\bibinfo{journal}{Nat. Photonics}}
  \textbf{\bibinfo{volume}{8}}, \bibinfo{pages}{145--152}
  (\bibinfo{year}{2014}).

\bibitem{GIPXTP5I}
\bibinfo{author}{Lucas, E.}, \bibinfo{author}{Guo, H.}, \bibinfo{author}{Jost,
  J.~D.}, \bibinfo{author}{Karpov, M.} \& \bibinfo{author}{Kippenberg, T.~J.}
\newblock \bibinfo{title}{Detuning-dependent properties and dispersion-induced
  instabilities of temporal dissipative kerr solitons in optical
  microresonators}.
\newblock \emph{\bibinfo{journal}{Phys. Rev. A}} \textbf{\bibinfo{volume}{95}},
  \bibinfo{pages}{043822} (\bibinfo{year}{2017}).

\bibitem{ERK7JBEB}
\bibinfo{author}{Li, X.} \emph{et~al.}
\newblock \bibinfo{title}{Universal isocontours for dissipative {Kerr}
  solitons}.
\newblock \emph{\bibinfo{journal}{Opt. Lett.}} \textbf{\bibinfo{volume}{43}},
  \bibinfo{pages}{2567} (\bibinfo{year}{2018}).

\bibitem{I28LWZJJ}
\bibinfo{author}{Coen, S.}, \bibinfo{author}{Randle, H.~G.},
  \bibinfo{author}{Sylvestre, T.} \& \bibinfo{author}{Erkintalo, M.}
\newblock \bibinfo{title}{Modeling of octave-spanning {Kerr} frequency combs
  using a generalized mean-field lugiato--lefever model}.
\newblock \emph{\bibinfo{journal}{Opt. Lett.}} \textbf{\bibinfo{volume}{38}},
  \bibinfo{pages}{37--39} (\bibinfo{year}{2013}).

\bibitem{5T6P52GN}
\bibinfo{author}{Karpov, M.} \emph{et~al.}
\newblock \bibinfo{title}{Raman self-frequency shift of dissipative {Kerr}
  solitons in an optical microresonator}.
\newblock \emph{\bibinfo{journal}{Phys. Rev. Lett.}}
  \textbf{\bibinfo{volume}{116}}, \bibinfo{pages}{103902}
  (\bibinfo{year}{2016}).

\bibitem{PSAFVI8W}
\bibinfo{author}{Yi, X.} \emph{et~al.}
\newblock \bibinfo{title}{Single-mode dispersive waves and soliton microcomb
  dynamics}.
\newblock \emph{\bibinfo{journal}{Nat. Commun.}} \textbf{\bibinfo{volume}{8}},
  \bibinfo{pages}{14869} (\bibinfo{year}{2017}).

\end{thebibliography}

\end{document}

\title{Supplemental Material for ``Power--Bandwidth Scaling of Resonantly Coupled Soliton Microcombs''}

\author{Xinrui Luo$^{1,*}$, Kaixuan Zhu$^{1,*}$, Yuanlei Wang$^{1,2*}$, Yinke Cheng$^{1,2}$, Haoyang Luo$^{1}$, Junqi Wang$^{1}$, Yiwen Yang$^{1}$, Zhenyu Xie$^{1}$, Bei-Bei Li$^{2}$, Qihuang Gong$^{1,3,4,5}$, and Qi-Fan Yang$^{1,3,4,5\dagger}$\\
$^1$State Key Laboratory for Artificial Microstructure and Mesoscopic Physics and Frontiers Science Center for Nano-optoelectronics, School of Physics, Peking University, Beijing 100871, China\\
$^2$Beijing National Laboratory for Condensed Matter Physics, Institute of Physics, Chinese Academy of Sciences, Beijing 100190, China\\
$^3$Peking University Yangtze Delta Institute of Optoelectronics, Nantong 226010, China\\
$^4$Collaborative Innovation Center of Extreme Optics, Shanxi University, Taiyuan 030006, China\\
$^5$Hefei National Laboratory, Hefei 230088, China\\
$^{*}$These authors contributed equally to this work.\\
$^{\dagger}$Corresponding author: leonardoyoung@pku.edu.cn}

%\date{\today}

\maketitle

\tableofcontents

\newpage

\section{Full theoretical model}
\label{sec:full-model}

\subsection{Coupled RC--NR equations}

We describe the system using the coupled resonant-coupler--nonlinear-resonator (RC--NR) model underlying the main text. The RC is represented by the resonance closest to the pump, while the NR retains the mode family required for comb formation. Coupling is restricted to the pumped modes. This model neglects the RC nonlinearity and assumes that neighboring RC resonances do not appreciably affect pump transfer or load the generated comb.

We use the normalization
\begin{equation}
\begin{gathered}
\tau=\frac{\kappa_\mathrm{NR}t}{2},
\qquad
\zeta_j=\frac{2(\omega_j-\omega_p)}{\kappa_\mathrm{NR}},
\qquad
j\in\{\mathrm{NR},\mathrm{RC}\},
\\
K=\frac{\kappa_\mathrm{RC}}{\kappa_\mathrm{NR}},
\qquad
K_e=\frac{\kappa_{e,\mathrm{RC}}}{\kappa_{e,\mathrm{NR}}},
\qquad
\mathcal G=\frac{2G}{\kappa_\mathrm{NR}},
\qquad
d_2=\frac{D_{2,\mathrm{NR}}}{\kappa_\mathrm{NR}}.
\end{gathered}
\label{eq:normalization}
\end{equation}
Here, $t$ is the physical slow time, $\kappa_j$ and $\kappa_{e,j}$ are the total decay and external coupling rates, $G$ is the inter-resonator coupling rate, and $\omega_j$ and $\omega_p$ are the cold-cavity resonance and pump angular frequencies. Both detunings are normalized to the NR decay rate. The NR second-order cavity-mode dispersion is $D_{2,\mathrm{NR}}$, with $d_2>0$ corresponding to anomalous dispersion under the convention used below.

The normalized fields obey \cite{JSUVFVVT,RNZK9EYV}
\begin{subequations}
\label{eq:coupled-model}
\begin{equation}
\frac{\partial\psi_\mathrm{RC}}{\partial\tau}
=
-(K+i\zeta_\mathrm{RC})\psi_\mathrm{RC}
+i\mathcal G\psi_{0,\mathrm{NR}}
+f_\mathrm{RC},
\label{eq:S2}
\end{equation}
\begin{equation}
\frac{\partial\psi_\mathrm{NR}}{\partial\tau}
=
-(1+i\zeta_\mathrm{NR})\psi_\mathrm{NR}
+id_2\frac{\partial^2\psi_\mathrm{NR}}{\partial\phi^2}
+i|\psi_\mathrm{NR}|^2\psi_\mathrm{NR}
+i\mathcal G\psi_\mathrm{RC}.
\label{eq:S3}
\end{equation}
\end{subequations}
Here, $\psi_\mathrm{RC}$ is the amplitude of the pumped RC mode, $\psi_\mathrm{NR}(\phi,\tau)$ is the NR field, and $\phi\in[-\pi,\pi)$ is the azimuthal coordinate. The NR modal amplitudes are defined by
\begin{equation}
\psi_\mathrm{NR}
=
\sum_\mu\psi_{\mu,\mathrm{NR}}e^{i\mu\phi},
\qquad
\psi_{0,\mathrm{NR}}
=
\frac{1}{2\pi}
\int_{-\pi}^{\pi}\psi_\mathrm{NR}\,d\phi.
\label{eq:S4}
\end{equation}
The spatially uniform RC field therefore drives only the pumped NR mode, $\mu=0$.

The applied pump amplitude $f_\mathrm{RC}$ is related to the bus-waveguide input power by
\begin{equation}
|f_\mathrm{RC}|^2
=
K_e\frac{P_\mathrm{in}}{P_0},
\qquad
P_0=
\frac{\hbar\omega_0\kappa_\mathrm{NR}^3}
{8g_\mathrm{NR}\kappa_{e,\mathrm{NR}}},
\label{eq:S8}
\end{equation}
where $\hbar$ is the reduced Planck constant, $\omega_0$ is the optical angular frequency, and $g_\mathrm{NR}$ is the NR Kerr nonlinear coupling coefficient. The same power scale $P_0$ is used for direct waveguide pumping.

\subsection{Effective pump power}

The effective drive delivered to the NR and its enhancement relative to direct pumping are defined by
\begin{equation}
f_\mathrm{eff}=i\mathcal G\psi_\mathrm{RC},
\qquad
|f_\mathrm{eff}|^2
=
\Gamma\frac{P_\mathrm{in}}{P_0}.
\label{eq:S9}
\end{equation}
Although the first relation is exact within the coupled model, the enhancement generally depends on the nonlinear NR state through the field returned to the RC.

Projecting Eq.~\eqref{eq:S3} onto the pumped mode gives
\begin{equation}
\frac{\partial\psi_{0,\mathrm{NR}}}{\partial\tau}
=
-(1+i\zeta_\mathrm{NR})\psi_{0,\mathrm{NR}}
+i\mathcal N_0
+f_\mathrm{eff},
\qquad
\mathcal N_0
=
\sum_{\mu,\nu}
\psi_{\mu,\mathrm{NR}}
\psi_{\nu,\mathrm{NR}}
\psi_{\mu+\nu,\mathrm{NR}}^*.
\label{eq:S11}
\end{equation}
To obtain the effective-pump curves used in the main text, we first omit $\mathcal N_0$ when calculating pump transfer. The Kerr nonlinearity is retained in the NR LLE describing comb formation and soliton dynamics. This approximation separates the linear resonant enhancement from the subsequent nonlinear response.

In steady state, the resulting pump-transfer relations are
\begin{equation}
\begin{aligned}
\psi_{0,\mathrm{NR}}
&=
\frac{i\mathcal G}{1+i\zeta_\mathrm{NR}}
\psi_\mathrm{RC},
\\
\psi_\mathrm{RC}
&=
\frac{f_\mathrm{RC}}
{
K+\dfrac{\mathcal G^2}{1+\zeta_\mathrm{NR}^2}
+i\left[
\zeta_\mathrm{RC}
-\dfrac{\mathcal G^2\zeta_\mathrm{NR}}
{1+\zeta_\mathrm{NR}^2}
\right]
}.
\end{aligned}
\label{eq:S15}
\end{equation}
The NR thus introduces an additional RC loading,
$\mathcal G^2/(1+\zeta_\mathrm{NR}^2)$, and a resonance shift,
$\mathcal G^2\zeta_\mathrm{NR}/(1+\zeta_\mathrm{NR}^2)$.
Using Eqs.~\eqref{eq:S8} and \eqref{eq:S9} gives
\begin{equation}
\Gamma(\zeta_\mathrm{NR},\zeta_\mathrm{RC})
=
\frac{K_e\mathcal G^2}
{
\left(
K+\dfrac{\mathcal G^2}{1+\zeta_\mathrm{NR}^2}
\right)^2
+
\left(
\zeta_\mathrm{RC}
-\dfrac{\mathcal G^2\zeta_\mathrm{NR}}
{1+\zeta_\mathrm{NR}^2}
\right)^2
}.
\label{eq:S19}
\end{equation}
For a given NR detuning, compensating the resonance shift maximizes the enhancement:
\begin{equation}
\zeta_\mathrm{RC}^{\mathrm{opt}}
=
\frac{\mathcal G^2\zeta_\mathrm{NR}}
{1+\zeta_\mathrm{NR}^2},
\qquad
\Gamma_\mathrm{max}(\zeta_\mathrm{NR})
=
\frac{K_e\mathcal G^2}
{
\left[
K+\mathcal G^2/(1+\zeta_\mathrm{NR}^2)
\right]^2
}.
\label{eq:S21}
\end{equation}

\begin{figure}[ht]
\centering
\includegraphics[width=0.98\linewidth]{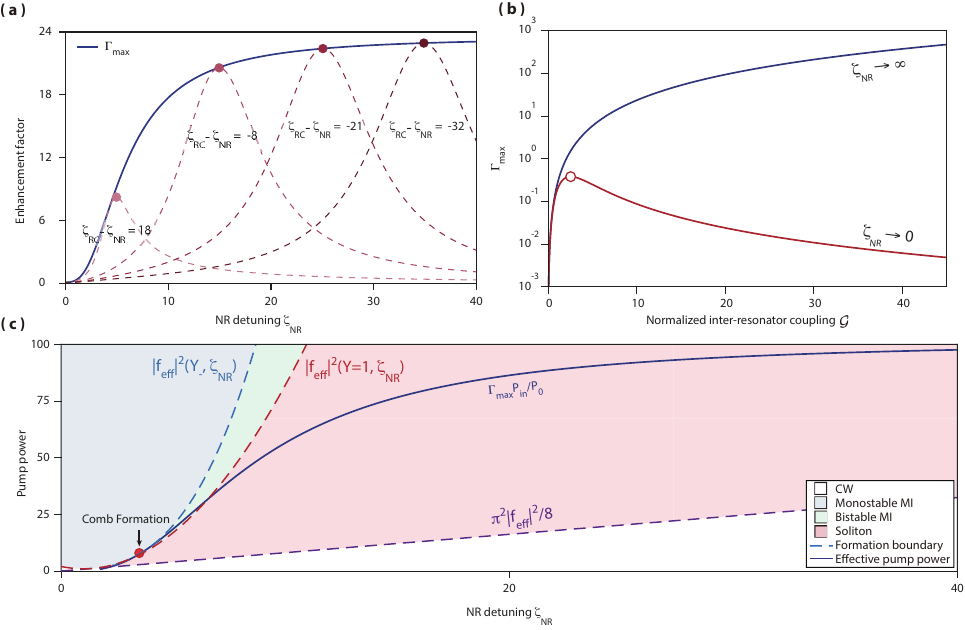}
\caption{\label{fig:S1}
\textbf{Effective pump power and soliton-formation boundary.}
\textbf{(a)} Pump-enhancement factor for four fixed RC--NR resonance offsets. The black curve gives $\Gamma_\mathrm{max}$, and circles mark contact with the optimized envelope.
\textbf{(b)} Maximum enhancement at zero NR detuning and its large-detuning limit $\Gamma_\infty$.
\textbf{(c)} CW MI-onset condition (red), high-input CW turning-point curve used to estimate the formation boundary (blue), and maximum effective pump power at fixed input power (gray dashed). The illustrated effective-pump curve uses the linear cavity response; the formation-power calculation includes the Kerr correction described in the text. Intersections identify candidate formation points, while soliton access is assessed along the corresponding tuning trajectory. The purple curve gives the ideal soliton sustainment bound \cite{GQ4W33CB}.}
\end{figure}

This maximum-enhancement trajectory is an envelope obtained by optimizing the RC resonance at each NR detuning. It is not generally followed during a laser scan with fixed cavity resonance frequencies, for which
$\zeta_\mathrm{RC}-\zeta_\mathrm{NR}
=2(\omega_\mathrm{RC}-\omega_\mathrm{NR})/\kappa_\mathrm{NR}$
remains constant. Repeating scans at different preset RC resonances samples different points on the envelope [Fig.~\ref{fig:S1}(a)], whereas following it continuously requires coordinated tuning of the RC resonance and pump. Applying the stationary transfer relation during tuning also assumes that the cavity fields remain close to their instantaneous stationary response.

At fixed coupling, the additional loading vanishes at large NR detuning, giving
\begin{equation}
\Gamma_\infty
=
\lim_{\zeta_\mathrm{NR}\rightarrow\infty}
\Gamma_\mathrm{max}(\zeta_\mathrm{NR})
=
K_e\frac{\mathcal G^2}{K^2}.
\label{eq:S22}
\end{equation}
Figure~\ref{fig:S1}(b) compares this asymptotic enhancement with the on-resonance value,
$\Gamma_\mathrm{max}(0)=K_e\mathcal G^2/(K+\mathcal G^2)^2$.
Whereas $\Gamma_\infty$ increases with coupling, the on-resonance enhancement decreases for $\mathcal G^2>K$, because the increased NR-induced loading outweighs the improvement in pump injection.

\subsection{Kerr correction to the pump transfer}

At the smaller detunings relevant to formation, the nonlinear shift of the pumped NR mode can appreciably modify the pump transfer. We separate its nonlinear contribution into
\begin{equation}
\mathcal N_0
=
\Delta_\mathrm{K}\psi_{0,\mathrm{NR}}
+
\mathcal N_0^{(\mathrm{FWM})},
\qquad
\Delta_\mathrm{K}
=
|\psi_{0,\mathrm{NR}}|^2
+
2\sum_{\mu\neq0}|\psi_{\mu,\mathrm{NR}}|^2.
\label{eq:S23}
\end{equation}
The first term accounts for self- and cross-phase modulation, while $\mathcal N_0^{(\mathrm{FWM})}$ contains the remaining phase-sensitive four-wave-mixing terms.

Retaining $\Delta_\mathrm{K}$ but omitting $\mathcal N_0^{(\mathrm{FWM})}$ in the stationary pump-transfer calculation gives
\begin{equation}
\psi_{0,\mathrm{NR}}
\simeq
\frac{f_\mathrm{eff}}
{1+i(\zeta_\mathrm{NR}-\Delta_\mathrm{K})}.
\label{eq:S25}
\end{equation}
The Kerr correction is therefore included by replacing
$\zeta_\mathrm{NR}$ with $\zeta_\mathrm{NR}-\Delta_\mathrm{K}$
in Eqs.~\eqref{eq:S19} and \eqref{eq:S21}.

For a continuous-wave (CW) field, this treatment is exact:
$\Delta_\mathrm{K}=|\psi_{0,\mathrm{NR}}|^2$ and
$\mathcal N_0^{(\mathrm{FWM})}=0$.
Once sidebands develop, it provides an estimate of the pump transfer rather than a complete description of the nonlinear state. The full coupled-LLE simulations retain all contributions to $\mathcal N_0$ and their self-consistent effect on both resonators.

\section{Input pump power required for soliton formation}
\label{sec:formation}

As the pump laser is scanned from the blue to the red side of the NR resonance, the initially CW field develops modulation instability (MI), whose sidebands can seed soliton formation. Following the main text, the MI region continuously accessible from the CW state is referred to as monostable MI. We estimate the formation power from the minimum input power required to reach its right boundary, using the stationary CW response to construct an analytical reference. Subsequent access to the soliton state is tested separately with the full coupled-LLE model.

\subsection{Formation boundary and minimum input power}

Writing the CW intensity as $Y=|\psi_{0,\mathrm{NR}}|^2$, the stationary NR response satisfies \cite{KX4LXUPF,76NN4HTK}
\begin{equation}
|f_\mathrm{eff}|^2
=
Y\left[1+(\zeta_\mathrm{NR}-Y)^2\right].
\label{eq:S32}
\end{equation}
At fixed detuning, its turning points obey
\begin{equation}
\frac{\partial|f_\mathrm{eff}|^2}{\partial Y}
=
3Y^2-4\zeta_\mathrm{NR}Y+1+\zeta_\mathrm{NR}^2
=
0.
\label{eq:S34}
\end{equation}
The lower-intensity root,
\begin{equation}
Y_-
=
\frac{
2\zeta_\mathrm{NR}-\sqrt{\zeta_\mathrm{NR}^2-3}
}{3},
\qquad
\zeta_\mathrm{NR}\geq\sqrt{3},
\label{eq:S35}
\end{equation}
corresponds to the high-input turning point. Substituting $Y_-$ into Eq.~\eqref{eq:S32} gives the reference formation boundary [Fig.~\ref{fig:S1}(c)]. Here, $Y_-$ specifies the CW state used to construct the boundary; it is not assumed to equal the pumped-mode intensity of a developed MI state.

To obtain a closed expression for the formation power, we evaluate the pump transfer using the Kerr shift of this CW reference, $\Delta_\mathrm{K}=Y_-$. Introducing the Kerr-shifted detuning $x=\zeta_\mathrm{NR}-Y_-$, the turning-point relations become
\begin{equation}
\begin{gathered}
Y_-=\frac{1+x^2}{2x},
\qquad
\zeta_\mathrm{NR}=\frac{1+3x^2}{2x},
\qquad
x\geq\frac{1}{\sqrt{3}},
\\
|f_\mathrm{eff}|^2
=
\frac{(1+x^2)^2}{2x}.
\end{gathered}
\label{eq:formation-parametrization}
\end{equation}
Optimizing the RC resonance at each boundary point gives
$\zeta_\mathrm{RC}=\mathcal G^2x/(1+x^2)$.
Combining the corresponding enhancement with Eq.~\eqref{eq:formation-parametrization} yields
\begin{equation}
\frac{P_\mathrm{in}}{P_0}
=
\frac{
\left[K(1+x^2)+\mathcal G^2\right]^2
}{
2K_e\mathcal G^2x
}.
\label{eq:formation-input-x}
\end{equation}
Minimization with respect to $x$ gives
$3Kx^2=K+\mathcal G^2$,
which lies within the allowed range for all positive coupling strengths. The resulting formation condition is
\begin{equation}
\begin{aligned}
P_\mathrm{in}\geq P_\mathrm{f}
&=
\frac{8\sqrt{3}}{9K_e}
\frac{
\sqrt K\left(K+\mathcal G^2\right)^{3/2}
}{
\mathcal G^2
}P_0
\\
&=
\frac{8\sqrt{3}}{9}
\frac{
\left[1+(K/K_e)\Gamma_\infty\right]^{3/2}
}{
\Gamma_\infty
}P_0.
\end{aligned}
\label{eq:S41}
\end{equation}

The nonmonotonic coupling dependence reflects the competition between pump injection and additional RC loading. At weak coupling, increasing $\mathcal G$ improves pump injection and lowers the formation power. The minimum occurs at $\mathcal G^2=2K$. Beyond this point, the increased loading raises the required power and shifts the optimum toward larger Kerr-shifted detuning. The opposing formation and sustainment trends discussed in the main text concern this increasing branch of $P_\mathrm{f}$.

Equation~\eqref{eq:S41} is optimized over both the boundary point and the RC resonance. For a scan with fixed cavity resonance frequencies, the selected point must also be reachable along the associated constant-offset trajectory described in Sec.~\ref{sec:full-model}. The optimized envelope alone does not establish this accessibility.

\subsection{Correction from developed sidebands}

When the MI sidebands contribute appreciably to the Kerr shift, their effect can be estimated using $\Delta_\mathrm{K}$ in Eq.~\eqref{eq:S23}. A single update consists of evaluating this shift from an NR modal field calculated with linear pump transfer, inserting it into Eq.~\eqref{eq:S25}, and repeating the input-power minimization along the same reference boundary. In this calculation, the boundary is still specified by Eqs.~\eqref{eq:S32} and \eqref{eq:S35}; only the transfer from the bus waveguide to the NR is updated.

This procedure includes the sideband contribution to cross-phase modulation but does not recompute the NR field self-consistently. The full coupled-LLE simulations instead retain both the readjustment of the modal field and the phase-sensitive four-wave-mixing terms, allowing the analytical estimate to be compared with the nonlinear dynamics.

\section{Input pump power required for soliton sustainment}
\label{sec:sustainment}

After formation, the NR detuning is increased to broaden the soliton spectrum, while the RC resonance is adjusted to enhance pump delivery. We first obtain the effective pump power required by the soliton energy balance and then convert it to the corresponding bus-waveguide input power.

\subsection{Effective-pump requirement}

For anomalous second-order dispersion and sufficiently large detuning, the single-soliton field is approximated by
\begin{equation}
\psi_\mathrm{NR}(\phi)
\simeq
\psi_\mathrm{cw}
+
\sqrt{2\zeta_\mathrm{NR}}\,
\operatorname{sech}
\left(
\sqrt{\frac{\zeta_\mathrm{NR}}{d_2}}\phi
\right)e^{i\theta},
\label{eq:S42}
\end{equation}
where $\psi_\mathrm{cw}$ is the weak CW background and $\theta$ is the soliton phase relative to the effective pump, whose phase is chosen to be zero. The pulse is assumed to be narrow compared with the resonator round trip, with the CW background contributing little to its energy balance.

Multiplying the stationary NR LLE by $\psi_\mathrm{NR}^*$, integrating over $\phi$, and taking the real part gives
\begin{equation}
\int_{-\pi}^{\pi}|\psi_\mathrm{NR}|^2\,d\phi
=
\operatorname{Re}
\left[
f_\mathrm{eff}
\int_{-\pi}^{\pi}\psi_\mathrm{NR}^*\,d\phi
\right].
\label{eq:S43}
\end{equation}
Substituting the soliton profile and neglecting the CW background yields
\begin{equation}
\cos\theta
=
\frac{\sqrt{8\zeta_\mathrm{NR}}}
{\pi|f_\mathrm{eff}|}.
\label{eq:S45}
\end{equation}
A stationary phase therefore requires \cite{GQ4W33CB,GIPXTP5I,ERK7JBEB}
\begin{equation}
|f_\mathrm{eff}|^2
\geq
\frac{8\zeta_\mathrm{NR}}{\pi^2}.
\label{eq:S46}
\end{equation}
This ideal sustainment bound follows from the soliton energy balance and is not, by itself, a complete stability condition.

\subsection{Conversion to input pump power}

For a specified stationary NR state, the RC equation gives the exact relation
\begin{equation}
f_\mathrm{RC}
=
\left[
K+i\zeta_\mathrm{RC}
+
\mathcal G^2
\frac{\psi_{0,\mathrm{NR}}}{f_\mathrm{eff}}
\right]\psi_\mathrm{RC}.
\label{eq:S47}
\end{equation}
The real and imaginary parts of
$\psi_{0,\mathrm{NR}}/f_\mathrm{eff}$
describe the additional RC loading and resonance shift produced by the NR, including its full nonlinear response. Optimizing the RC detuning for this state gives
\begin{equation}
\begin{aligned}
\zeta_\mathrm{RC}^{\mathrm{opt}}
&=
-\mathcal G^2
\operatorname{Im}
\left(
\frac{\psi_{0,\mathrm{NR}}}{f_\mathrm{eff}}
\right),
\\
\Gamma_\mathrm{max}
&=
\frac{K_e\mathcal G^2}
{
\left[
K+
\mathcal G^2
\operatorname{Re}
\left(
\frac{\psi_{0,\mathrm{NR}}}{f_\mathrm{eff}}
\right)
\right]^2
}.
\end{aligned}
\label{eq:S49}
\end{equation}
Equation~\eqref{eq:S43} implies
$\operatorname{Re}(\psi_{0,\mathrm{NR}}/f_\mathrm{eff})\geq0$,
so the stationary NR adds nonnegative loading and
$\Gamma_\mathrm{max}\leq\Gamma_\infty$.
Approaching the asymptotic enhancement therefore requires this loading to be small compared with $K$.

Near the ideal sustainment bound, the pumped Fourier component of Eq.~\eqref{eq:S42} is
\begin{equation}
\psi_{0,\mathrm{NR}}
\simeq
\sqrt{\frac{d_2}{2}}e^{i\theta}
+
O\!\left(\zeta_\mathrm{NR}^{-1/2}\right),
\label{eq:S50}
\end{equation}
where the correction includes the CW background. The soliton contribution approaches a constant, whereas
$|f_\mathrm{eff}|\propto\sqrt{\zeta_\mathrm{NR}}$.
At fixed coupling, the relative effect of the NR on the RC consequently decreases with detuning.

A sufficient condition for both the additional loading and resonance shift to be small is
\begin{equation}
\frac{\mathcal G^2}{K}
\left|
\frac{\psi_{0,\mathrm{NR}}}{f_\mathrm{eff}}
\right|
\ll1.
\label{eq:S52}
\end{equation}
Then
$|\zeta_\mathrm{RC}^{\mathrm{opt}}|\ll K$
and
$\Gamma_\mathrm{max}\simeq\Gamma_\infty$.
This condition concerns the pump transfer, not the importance of the Kerr nonlinearity to the soliton itself.

Combining the ideal sustainment bound with the asymptotic enhancement gives
\begin{equation}
P_\mathrm{in}
\geq
\frac{8\zeta_\mathrm{NR}}{\pi^2\Gamma_\infty}P_0
=
\frac{8\zeta_\mathrm{NR}K^2}
{\pi^2K_e\mathcal G^2}P_0.
\label{eq:S55}
\end{equation}
This is the sustainment condition used in the main text. Because
$\Gamma_\mathrm{max}\leq\Gamma_\infty$,
it remains a lower bound within the soliton approximation. Using it to estimate the actual sustainment power additionally requires Eq.~\eqref{eq:S52}; otherwise, the state-dependent enhancement in Eq.~\eqref{eq:S49} must be retained.

\subsection{Optimal coupling and power--bandwidth scaling}

For an input power held fixed throughout formation and subsequent tuning, the minimum required power is set by the larger of the two requirements. Denoting the right-hand side of Eq.~\eqref{eq:S55} by $P_\mathrm{s}$, the analytical optimization at fixed $K$, $K_e$, and NR parameters is
\begin{equation}
P_\mathrm{in,min}
=
\min_{\mathcal G}
\max\left[
P_\mathrm{f}(\mathcal G),
P_\mathrm{s}(\mathcal G;\zeta_\mathrm{NR})
\right].
\label{eq:power-minimax}
\end{equation}
On the increasing branch of the formation requirement, the minimum occurs at the intersection of the two power curves. Equating Eqs.~\eqref{eq:S41} and \eqref{eq:S55} gives
\begin{equation}
\mathcal G_\mathrm{opt}^2
=
K\left(
\frac{3\zeta_\mathrm{NR}^{2/3}}{\pi^{4/3}}-1
\right).
\label{eq:S56}
\end{equation}
The intersection lies on this branch,
$\mathcal G_\mathrm{opt}^2\geq2K$,
for $\zeta_\mathrm{NR}\geq\pi^2$, as appropriate to the large-detuning analysis.

The resulting minimum input power is
\begin{equation}
\frac{P_\mathrm{in,min}}{P_0}
=
\frac{
8K\zeta_\mathrm{NR}
}{
\pi^2K_e
\left(
3\zeta_\mathrm{NR}^{2/3}/\pi^{4/3}-1
\right)
}
\approx
\frac{
8K\zeta_\mathrm{NR}^{1/3}
}{
3K_e\pi^{2/3}
},
\label{eq:S57}
\end{equation}
where the final approximation applies at large detuning. For the soliton profile in Eq.~\eqref{eq:S42}, the 3-dB bandwidth satisfies
$B_\mathrm{3dB}\propto\sqrt{\zeta_\mathrm{NR}}$
at fixed dispersion and free spectral range \cite{JSUVFVVT}. Thus,
\begin{equation}
P_\mathrm{in,min}\propto B_\mathrm{3dB}^{2/3},
\qquad
P_\mathrm{in}^{(\mathrm{direct})}\propto B_\mathrm{3dB}^{2},
\label{eq:bandwidth-scaling}
\end{equation}
with the direct-pumping relation following from Eq.~\eqref{eq:S46}.

The resonant-coupling scaling describes a family of designs with the coupling optimized for each target bandwidth, rather than a bandwidth sweep at fixed coupling. The small-loading condition in Eq.~\eqref{eq:S52} must therefore be checked along this optimized family: increasing detuning alone does not ensure negligible loading when $\mathcal G$ also increases. The predicted scaling applies where this condition and the soliton approximation hold together.

\section{Numerical simulations}
\label{sec:numerics}

We solve Eqs.~\eqref{eq:S2} and \eqref{eq:S3} using the split-step Fourier method, using 1024 NR modes with $d_2 = 1.2\times10^{-2}$. The simulations retain the RC dynamics and the full Kerr response of the NR.

\subsection{Simulation of formation power}

The numerical formation-power estimate is obtained from the evolution of weak sidebands around an initial CW state. The total sideband amplitude is
$A_\mathrm{sb}(\tau)
=[\sum_{\mu\neq0}|\psi_{\mu,\mathrm{NR}}(\tau)|^2]^{1/2}$,
with corresponding power $P_\mathrm{sb}=A_\mathrm{sb}^2$.
We fit
$\ln[A_\mathrm{sb}(\tau)/A_\mathrm{sb}(\tau_0)]$
to a linear function of $\tau$ over an interval that excludes the initial CW transient and subsequent nonlinear saturation. The fitted slope $\sigma_\mathrm{fit}$ is therefore the small-signal amplitude growth rate.

Positive and negative values of $\sigma_\mathrm{fit}$ indicate sideband growth and decay, respectively, while zero growth identifies the MI-onset boundary. The corresponding power evolves as
$P_\mathrm{sb}\propto\exp(2\sigma_\mathrm{fit}\tau)$.
Representative traces are shown in the lower panel of Fig.~\ref{fig:S2}(a).

The growth map in Fig.~\ref{fig:S2}(a) is calculated at
$\Gamma_\infty=50$, $K=38.5$, and $K_e=88.2$.
Each simulation starts from the corresponding CW state with a weak broadband perturbation. A preceding scan over NR detuning, RC detuning, and input power identifies the low-power instability region. The pump-relative RC detuning is then set to $\zeta_\mathrm{RC}=260$ and held fixed for the two-dimensional map. This numerical constraint differs from holding the physical RC resonance fixed during a laser scan, which instead keeps
$\zeta_\mathrm{RC}-\zeta_\mathrm{NR}$
constant.

Linear interpolation between adjacent points with positive and negative growth rates gives the zero-growth contour. Its lowest point defines the instability threshold at the scan resolution. Repeating the procedure over the coupling strengths considered gives the numerical formation-power estimates in Fig.~2(a) of the main text.

We test soliton access separately by first evolving the system at fixed detuning near the formation threshold until sidebands develop. The NR detuning is then increased at fixed input power, while the RC detuning is adjusted to maximize pump enhancement. Figure~\ref{fig:S2}(b) shows a representative evolution from CW through MI to a single soliton at a final detuning of $\zeta_\mathrm{NR}=153$.

\subsection{Simulation of sustainment power}

For the experimental devices, the sustainment calculation additionally includes the measured integrated dispersion and the Raman response \cite{I28LWZJJ,5T6P52GN,PSAFVI8W}. We write this model in physical units using
\begin{equation}
\widetilde A_\mu
=
\sqrt{\frac{\kappa_\mathrm{NR}}{2g_\mathrm{NR}}}
\,\psi_{\mu,\mathrm{NR}},
\qquad
B_0
=
\sqrt{\frac{\kappa_\mathrm{NR}}{2g_\mathrm{NR}}}
\,\psi_\mathrm{RC},
\label{eq:dimensional-fields}
\end{equation}
whose squared moduli give the photon numbers in the corresponding modes. The NR fast-time field is denoted by $A(t,t_\mathrm{f})$, where $t_\mathrm{f}$ is the coordinate within one round trip. The Fourier normalization is chosen such that the round-trip average of $|A|^2$ equals
$\sum_\mu|\widetilde A_\mu|^2$.

\begin{figure}
\centering
\includegraphics[width=0.98\linewidth]{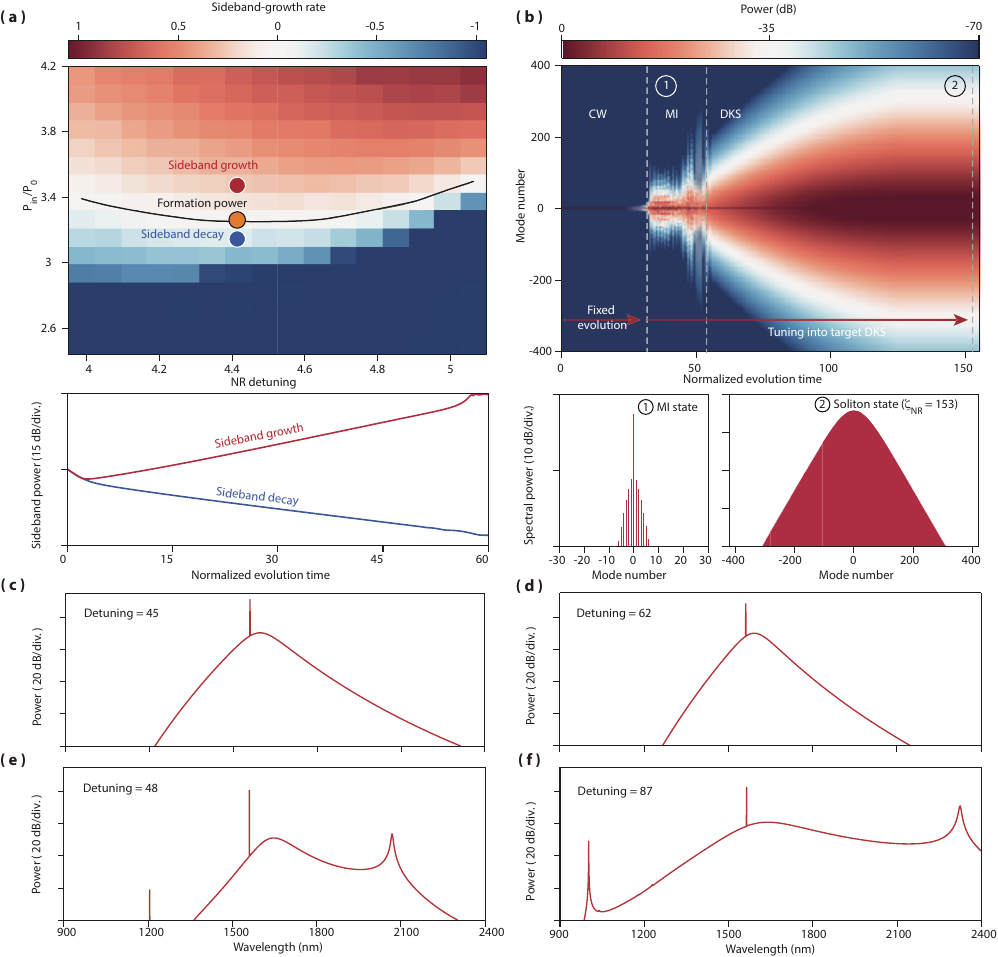}
\caption{\label{fig:S2}
\textbf{Numerical simulations of soliton formation and sustainment.}
\textbf{(a)} Top: fitted sideband growth rate $\sigma_\mathrm{fit}$ versus NR detuning and normalized input power at $\Gamma_\infty=50$, starting from weakly perturbed CW states. The RC detuning is selected by a preceding optimization and held fixed during the scan. The black contour marks zero growth, and the gold marker identifies its lowest-power point. Bottom: representative sideband-power traces showing growth and decay.
\textbf{(b)} Top: fixed-input-power evolution from CW through MI to a single soliton. After an initial fixed-detuning interval, the NR detuning is increased while the RC detuning is adjusted to maximize pump enhancement. Bottom: spectra at the marked MI and soliton stages. The final detuning is $\zeta_\mathrm{NR}=153$.
\textbf{(c)} Target single-soliton spectrum for a 100-GHz-FSR device
at $\zeta_\mathrm{NR}=45$, corresponding to the sustainment curve
in Fig.~3(d) of the main text.
\textbf{(d)} Target single-soliton spectrum for a 100-GHz-FSR device
at $\zeta_\mathrm{NR}=62$, corresponding to the other sustainment curve
in Fig.~3(d) of the main text.
\textbf{(e)} Target single-soliton spectrum for the 25-GHz-FSR device
at $\zeta_\mathrm{NR}=48$, corresponding to the sustainment curve
in the left inset of Fig.~4(a) of the main text.
\textbf{(f)} Target single-soliton spectrum for the 100-GHz-FSR device
at $\zeta_\mathrm{NR}=87$, corresponding to the sustainment curve
in the left inset of Fig.~4(b) of the main text.}
\end{figure}

Using the same detunings and decay-rate ratio as in Eq.~\eqref{eq:normalization}, the dimensional equations are
\begin{equation}
\begin{aligned}
\frac{\partial A}{\partial t}
={}&
-\frac{\kappa_\mathrm{NR}}{2}
(1+i\zeta_\mathrm{NR})A
-i\mathcal F^{-1}
\left[
D_{\mathrm{int,NR}}(\mu)\widetilde A_\mu
\right]
\\
&+iGB_0
+ig_\mathrm{NR}|A|^2A
-ig_\mathrm{NR}\tau_R A
\frac{\partial|A|^2}{\partial t_\mathrm{f}},
\end{aligned}
\label{eq:S58}
\end{equation}
and
\begin{equation}
\frac{dB_0}{dt}
=
-\frac{\kappa_\mathrm{NR}}{2}
(K+i\zeta_\mathrm{RC})B_0
+iG\widetilde A_0
+
\sqrt{
\frac{\kappa_{e,\mathrm{RC}}P_\mathrm{in}}
{\hbar\omega_0}
}.
\label{eq:S59}
\end{equation}
The integrated dispersion is
\begin{equation}
D_{\mathrm{int,NR}}(\mu)
=
\omega_{\mu,\mathrm{NR}}
-\omega_\mathrm{NR}
-\mu D_{1,\mathrm{NR}},
\label{eq:S61}
\end{equation}
where $\omega_{\mu,\mathrm{NR}}$ is the frequency of NR mode $\mu$ and
$D_{1,\mathrm{NR}}/(2\pi)$ is the free spectral range.
The operator $\mathcal F^{-1}$ transforms the modal dispersion term to the fast-time domain. Retaining only
$D_{\mathrm{int,NR}}=D_{2,\mathrm{NR}}\mu^2/2$
recovers the dispersion term in Eq.~\eqref{eq:S3} after normalization.

The final term in Eq.~\eqref{eq:S58} represents the delayed Raman response through its first temporal moment, characterized by the effective Raman shock time $\tau_R$ \cite{5T6P52GN}. This approximation includes the Raman contribution to the soliton self-frequency shift.

For each device, we first determine the minimum direct-pumping power
$P_\mathrm{s}^{(\mathrm{direct})}$
that sustains the target single-soliton state. This calculation omits the RC equation and replaces $iGB_0$ in Eq.~\eqref{eq:S58} with
$\sqrt{\kappa_{e,\mathrm{NR}}
P_\mathrm{s}^{(\mathrm{direct})}/(\hbar\omega_0)}$.
It therefore isolates the sustainment requirement associated with the measured dispersion and Raman response from the resonant pump enhancement.

For the same target NR state under resonant pumping, the corresponding bus-waveguide input power is
\begin{equation}
\frac{P_\mathrm{s}}{P_0}
=
\frac{
P_\mathrm{s}^{(\mathrm{direct})}/P_0
}{
\Gamma_\mathrm{max}
}
\simeq
\frac{
P_\mathrm{s}^{(\mathrm{direct})}/P_0
}{
\Gamma_\infty
},
\label{eq:S64}
\end{equation}
where the approximation uses the small-loading condition in Eq.~\eqref{eq:S52}. The descending sustainment curves are thus obtained from the generalized-LLE direct-pumping results and the resonant enhancement, rather than from the ideal second-order-dispersion bound alone.

The target single-soliton spectra are shown in Figs.~\ref{fig:S2}(c)--(f). Their normalized NR detunings are $\zeta_\mathrm{NR}=45$, $62$, $48$, and $87$, with corresponding minimum normalized direct-pumping sustainment powers $P_\mathrm{s}^{(\mathrm{direct})}/P_0=667$, $1342$, $282$, and $1300$, respectively. The normalized second-order dispersions are $d_2=3.1\times10^{-3}$, $6.8\times10^{-3}$, $1.9\times10^{-4}$, and $2.3\times10^{-3}$ in the same order, with $\tau_R=0.4\,\mathrm{fs}$ throughout. Using Eq.~\eqref{eq:S64}, the first two sustainment powers generate the curves in Fig.~3(d) of the main text, while the latter two generate those in the left insets of Figs.~4(a) and 4(b), respectively. These curves capture the measured decrease in sustainment power with increasing $\Gamma_\infty$ over the sampled operating range. The corresponding RC parameters are $K=9.69$, $8.97$, $16.27$, and $6.31$ and $K_e=11.26$, $11.00$, $26.96$, and $8.29$, respectively, with $\mathcal{G}$ determined by $\Gamma_\infty$ for each parameter set.

%\clearpage

\bibliography{supplement}